\documentclass[sigconf]{acmart}
\AtBeginDocument{%
  \providecommand\BibTeX{{%
    \normalfont B\kern-0.5em{\scshape i\kern-0.25em b}\kern-0.8em\TeX}}}

\copyrightyear{2024} 
\acmYear{2024} 
\setcopyright{rightsretained} 
\acmConference[RecSys '24]{18th ACM Conference on Recommender Systems}{October 14--18, 2024}{Bari, Italy}
\acmBooktitle{18th ACM Conference on Recommender Systems (RecSys '24), October 14--18, 2024, Bari, Italy}
\acmDOI{10.1145/3640457.3688064}
\acmISBN{979-8-4007-0505-2/24/10}

\usepackage{bm}
\usepackage{enumitem}
\usepackage{makecell}
\usepackage{multirow}
\usepackage{bbding}
\usepackage{xcolor,soul}
\usepackage{balance}

\DeclareMathOperator{\degree}{deg}
\DeclareMathOperator{\dist}{dist}
\DeclareMathOperator{\timestamp}{time}
\DeclareMathOperator{\pagerank}{pagerank}

\definecolor{lightgray}{RGB}{225, 225, 225}

\begin{document}

\title{Fair Augmentation for Graph Collaborative Filtering}


\author{Ludovico Boratto}
\orcid{0000-0002-6053-3015}
\affiliation{%
  \institution{University of Cagliari}
  \streetaddress{Via Ospedale, 72}
  \city{Cagliari}
  \country{Italy}
}
\email{ludovico.boratto@acm.org}

\author{Francesco Fabbri}
\orcid{0000-0002-9631-1799}
\affiliation{%
  \institution{Spotify}
  \city{Barcelona}
  \country{Spain}
}
\email{francescof@spotify.com}

\author{Gianni Fenu}
\orcid{0000-0003-4668-2476}
\affiliation{%
  \institution{University of Cagliari}
  \streetaddress{Via Ospedale, 72}
  \city{Cagliari}
  \country{Italy}
}
\email{fenu@unica.it}

\author{Mirko Marras}
\orcid{0000-0003-1989-6057}
\affiliation{%
  \institution{University of Cagliari}
  \streetaddress{Via Ospedale, 72}
  \city{Cagliari}
  \country{Italy}
}
\email{mirko.marras@acm.org}

\author{Giacomo Medda}
\orcid{0000-0002-1300-1876}
\affiliation{%
  \institution{University of Cagliari}
  \streetaddress{Via Ospedale, 72}
  \city{Cagliari}
  \country{Italy}
}
\email{giacomo.medda@unica.it}

\renewcommand{\shortauthors}{Boratto et al.}

\begin{abstract}
  Recent developments in recommendation have harnessed the collaborative power of graph neural networks (GNNs) in learning users' preferences from user-item networks. Despite emerging regulations addressing fairness of automated systems, unfairness issues in graph collaborative filtering remain underexplored, especially from the consumer's perspective. Despite numerous contributions on consumer unfairness, only a few of these works have delved into GNNs. A notable gap exists in the formalization of the latest mitigation algorithms, as well as in their effectiveness and reliability on cutting-edge models. This paper serves as a solid response to recent research highlighting unfairness issues in graph collaborative filtering by reproducing one of the latest mitigation methods. The reproduced technique adjusts the system fairness level by learning a fair graph augmentation. Under an experimental setup based on 11 GNNs, 5 non-GNN models, and 5 real-world networks across diverse domains, our investigation reveals that fair graph augmentation is consistently effective on high-utility models and large datasets. Experiments on the transferability of the fair augmented graph open new issues for future recommendation studies. Source code: \url{https://github.com/jackmedda/FA4GCF}.
\end{abstract}




\maketitle

\section{Introduction}

Recommender systems are powerful technologies that automatically learn the users' preferences towards diverse categories of items, ranging from streams to points of interest.
The worldwide usage of recommender systems not only spotlights their usefulness, but also raises worrying doubts regarding their trustworthiness~\cite{WangZWLR22}, robustness~\cite{DBLP:journals/corr/abs-2309-02057}, explainability~\cite{ZhangC20}, and fairness~\cite{WanngMZLM22}.

Focusing on concerns about unfairness, the literature in recommendation has been particularly devoted to devising algorithms to mitigate the inequality exhibited by recommendations~\cite{DBLP:conf/recsys/TsintzouPT19}.
Researchers' efforts can be categorized based on the step a fairness-aware algorithm takes action~\cite{DBLP:journals/csur/MehrabiMSLG21}: before (\textit{pre-processing}), during (\textit{in-processing}), and after (\textit{post-processing}) the training phase.
{\color{black} Typically, pre-processing approaches involve manipulating the input data before feeding it into the models, in-processing approaches are driven by multi-objective optimization techniques including a fairness-aware function, and post-processing approaches focus on re-ranking the recommendations to satisfy the targeted fairness goal.}
Although pre- and post-processing methods offer benefits such as being mostly model-agnostic and not requiring re-training~\cite{DBLP:journals/ipm/BorattoFMM23,DBLP:conf/interspeech/FenuMMM21}, most of the works fall into the in-processing category, especially mitigation algorithms devised for consumer unfairness~\cite{DBLP:journals/air/VassoyL24,DBLP:conf/icb/AtzoriFM22,DBLP:journals/jstsp/AtzoriFM23}.

The issue is particularly notable in fairness-aware graph collaborative filtering (GCF)~\cite{DBLP:journals/air/VassoyL24}, where the recommendation task is optimized through (knowledge) graphs and graph neural networks (GNNs).
GCF has recently attracted strong attention in the research landscape thanks to i) the ability of GNNs to capture high-order connectivites between users and items~\cite{DBLP:conf/sigir/Wang0WFC19,DBLP:conf/recsys/AnelliMPBSN23}, and ii) the need to build a solid theoretical foundation about information flow in GNN-based recommenders~\cite{DBLP:conf/cikm/ShenWZSZLL21,DBLP:conf/cikm/MaoZXLWH21,DBLP:conf/cikm/PengSM22}.
Despite recent advancements in GCF, \cite{DBLP:journals/air/VassoyL24} has emphasized the insufficient exploration of data augmentation (pre-processing) and re-ranking (post-processing) procedures to counteract consumer unfairness.
{\color{black} The former technique aims to satisfy the fairness goal within a pre-processing stage, while the latter re-order the recommendations generated by graph-based systems within a post-processing phase.}
Several studies have showcased the benefits offered by data augmentation in GCF~\cite{DBLP:conf/sigir/LeeKJPY21,DBLP:conf/sigir/FanXDPZY23}, but fairness-aware works in GCF mostly address providers~\cite{DBLP:conf/um/MansouryAPMB20,DBLP:conf/cikm/LuoMXS23,DBLP:conf/www/ChenFCLLZL23}, or fall into the in-processing category~\cite{DBLP:conf/kdd/DongKTL21,DBLP:journals/tist/LiHZ22}.

The gap and the associated concerns emphasized by \cite{DBLP:journals/air/VassoyL24} have recently been investigated in GCF by means of a post-processing data augmentation pipeline~\cite{DBLP:conf/cikm/BorattoFFMM23}.
This method learns through a trained GNN how to augment a graph to provide fairer recommendations.
{\color{black} Their main contributions come from the optimization of a fairness-aware loss function that monitors the gap in recommendation utility across demographic groups by leveraging the relevance judgements of unseen interaction data, e.g., the validation set.}
Despite remarkable performance, the evaluation assessment was conducted on a limited set of datasets and models.
As illustrated by recent works~\cite{DBLP:conf/ecir/AnelliDNMPP23}, evaluation studies in GCF require a comprehensive protocol including a broad range of GNNs and datasets.
Another concern lies in the recency of the tested GNNs, namely GCMC~\cite{DBLP:journals/corr/BergKW17}, LightGCN~\cite{DBLP:conf/sigir/0001DWLZ020}, NGCF~\cite{DBLP:conf/sigir/Wang0WFC19}, which nowadays represent consolidated baselines in GCF~\cite{DBLP:conf/ecir/AnelliDNMPP23,DBLP:conf/recsys/AnelliMPBSN23}.
{\color{black} Some of these models, e.g., LightGCN, report state-of-the-art performance, but a large set of novel architectures for GCF have not been explored in their study~\cite{DBLP:conf/www/XiaHHLYK23,DBLP:conf/cikm/PengSM22,DBLP:conf/kdd/WangYM000M22,DBLP:conf/cikm/MaoZXLWH21,DBLP:journals/tkde/YuXCCHY24}.}
Furthermore, \cite{DBLP:conf/cikm/BorattoFFMM23} proposed a set of sampling policies to restrict the nodes targeted by the augmentation process.
However, such policies were only deduced from an empirical reasoning that overlooked relevant aspects.
Specifically, the policies are not backed up by a solid formalization of their functionality and their impact on the model architecture, and common interaction data features and classic graph properties were not considered to define the policies.

In this paper, we address the concerns regarding the discussed mitigation algorithm~\cite{DBLP:conf/cikm/BorattoFFMM23} by providing a mix of reproducibility and replicability study~\cite{acmbadging}, similarly to recent works~\cite{DBLP:conf/sigir/RahmaniNDA22,DBLP:conf/recsys/DacremaCJ19}.
Specifically, (i) we provide a theoretical formalization of the sampling policies and of the augmented graph integration in the GNNs forward process, (ii) we extend the set of sampling policies by considering the interaction time and classical graph properties, (iii) we conduct an exhaustive benchmark of consumer unfairness mitigation across age and gender groups over an evaluation protocol that includes 5 real-world networks, 11 GNN-based and 5 non-GNN recommender systems.
Furthermore, our work aims to provide a solid framework that extends the mitigation algorithm provided by \cite{DBLP:conf/cikm/BorattoFFMM23}, which adopted inefficient GNN backbones and a restricted set of graph architectures.
In detail, we publicly\footnote{\url{https://github.com/jackmedda/FA4GCF}} offer FA4GCF (\textbf{F}air \textbf{A}ugmentation for \textbf{G}raph \textbf{C}ollaborative \textbf{F}iltering), a resource built upon Recbole~\cite{DBLP:conf/cikm/ZhaoMHLCPLLWTMF21,DBLP:conf/cikm/ZhaoHPYZLZBTSCX22} that can be seamlessly adapted to any GNNs, datasets, sensitive attributes, or sampling policies.

\section{Reproducibility Study}

In this section, we describe the reproducibility methodology of the considered approach.
Particularly, we briefly introduce the notation of the recommendation task in GCF, we dive into the implementation of the reproduced method, and present our evaluation setup.

\subsection{GNN-based Recommendation Task}

In recommendation, models typically learn the users' preferences from past interactions between a user set $\mathcal{U}$ and an item set $\mathcal{I}$.
The interactions are represented by an implicit feedback matrix $R \in \{0, 1\}^{|\mathcal{U}| \times |\mathcal{I}|}$, where $R_{u,i} = 1$ denotes user $u$ interacted with item $i$.
The interactions can also be viewed as the edges of an undirected bipartite graph $\mathcal{G} = (\mathcal{U}, \mathcal{I}, E)$, where $\mathcal{U} \cup \mathcal{I}$ is the set of nodes, $E$ is the set of edges.
In GCF, GNNs are adapted to generate recommendations by solving a linking prediction task on the graph $\mathcal{G}$, which can be encoded by a symmetric adjacency matrix $A$:

\begin{equation}
    A = 
    \begin{gathered}
        \begin{bmatrix}
        0 & R\\
        R^T & 0
        \end{bmatrix}
    \end{gathered}
\end{equation}

Typically, a GNN $f$ predicts the missing entries in $R$ through message-passing schemes~\cite{DBLP:journals/tors/GaoZLLQPQCJHL23}.
The predicted linking probabilities $\hat{R}$ between users and items are sorted in descending order, then the top-$n$ items are recommended to the corresponding users.

\begin{table*}[]
    \centering
    \caption{Dataset statistics. \emph{M} and \emph{F} respectively stand for \emph{Males} and \emph{Females}, \emph{Y} and \emph{O} respectively stand for \emph{Younger} and \emph{Older}.}
    \resizebox{0.8\textwidth}{!}{
    \begin{tabular}{r|r|r|r|r|r}
    \toprule
    & FNYC~\cite{DBLP:journals/tist/YangZQ16,DBLP:conf/huc/YangZQC16} & FTKY~\cite{DBLP:journals/tist/YangZQ16,DBLP:conf/huc/YangZQC16} & LF1M~\cite{DBLP:conf/ecir/BalloccuBCFM23} & ML1M~\cite{DBLP:journals/tiis/HarperK16} & RENT~\cite{DBLP:conf/recsys/MisraWM18} \\
    \midrule
    \# Users & 4,832 & 7,240 & 4,546 & 6,040 & 5,050 \\
    \# Items & 5,651 & 5,785 & 12,492 & 3,706 & 3,397 \\
    \# Interactions & 182,164 & 353,847 & 1,082,132 & 1,000,209 & 43,770 \\
    Gender Representation & M : 62.8\%; F : 37.2\% & M : 87.9\%; F : 12.1\% & M : 78.5\%; F : 21.5\% & M : 71.7\%; F : 28.3\% & - \\
    Age Representation & - & - & Y : 60.8\%; O : 39.2\% & Y : 56.6\%; O : 43.4\% & Y : 53.8\%; O : 46.2\% \\
    Minimum Degree per user & 20 & 20 & 20 & 20 & 5 \\
    \bottomrule
    \end{tabular}
    }
    \label{tab:datasets}
\end{table*}

\begin{table*}
    \centering
    \caption{GNN-based models characteristics. \emph{E} and \emph{I} denote the \emph{Explicit} and \emph{Implicit} taxonomy proposed by \cite{DBLP:conf/ecir/AnelliDNMPP23} for the message-passing type of GNNs. \emph{Augmentable} refers to the capacity of a GNN-based system to be used with the augmentation process~\cite{DBLP:conf/cikm/BorattoFFMM23}.}
    \resizebox{\textwidth}{!}{
    \begin{tabular}{l|rrrrrrrrrrr}
    \toprule
     & AutoCF~\cite{DBLP:conf/www/XiaHHLYK23} & DirectAU~\cite{DBLP:conf/kdd/WangYM000M22}$^{*}$ & HMLET~\cite{DBLP:conf/wsdm/KongKJ0LPK22} & LightGCN~\cite{DBLP:conf/sigir/0001DWLZ020} & NCL~\cite{DBLP:conf/www/LinTHZ22} & NGCF~\cite{DBLP:conf/sigir/Wang0WFC19} & SGL~\cite{DBLP:conf/sigir/WuWF0CLX21} & SVD-GCN~\cite{DBLP:conf/cikm/PengSM22} & SVD-GCN-S~\cite{DBLP:conf/cikm/PengSM22} & UltraGCN~\cite{DBLP:conf/cikm/MaoZXLWH21} & XSimGCL~\cite{DBLP:journals/tkde/YuXCCHY24} \\
    \midrule
    Conference/Journal & WWW & KDD & WSDM & SIGIR & WWW & SIGIR & SIGIR & CIKM & CIKM & CIKM & TKDE \\
    Year & 2023 & 2022 & 2022 & 2020 & 2022 & 2019 & 2021 & 2022 & 2022 & 2021 & 2024 \\
    Message-passing & E & E & E & E & E & E & E & I & I & I & E \\
    Augmentable & \Checkmark & \Checkmark & \Checkmark & \Checkmark & \Checkmark & \Checkmark & \Checkmark & \Checkmark & & & \Checkmark \\
    \bottomrule
    \addlinespace[3pt]
    \multicolumn{12}{l}{\large $^{*}$LightGCN is adopted as the base encoder.}
    \end{tabular}
    }
    \label{tab:models}
\end{table*}

\subsection{Reproduced Approach}

\subsubsection{Graph Augmentation Pipeline} \label{subsec:graph-aug}

Briefly, \cite{DBLP:conf/cikm/BorattoFFMM23} proposed to lighten the edge-perturbation process of GNN explainers~\cite{DBLP:conf/aistats/LucicHTRS22,DBLP:conf/pkdd/KangLB21} and extend it to perturbing zero entries in the adjacency matrix $A$, essentially augmenting the graph.
The augmentation algorithm samples edges from a predefined set $\tilde{E}$ ($\tilde{E} \cap E = \emptyset$) of candidate edges, which includes at most all of the user-item interactions missing from $R$.
The subset of edges actually added are learnt by iteratively updating a perturbation vector $p$, which determines the addition of new edges based on a fixed positional relation.
This iterative pipeline results in the augmented graph $\tilde{A}$, which can be fed into the trained GNN $f$ to retrieve the altered and potentially fair feedback matrix~$\tilde{R}$.

The iterative process aims to optimize a loss function of the form:
\begin{equation} \label{eq:loss}
    \mathcal{L}(A, \tilde{A}) = \mathcal{L}_{fair}(A, \tilde{R}) + \beta \mathcal{L}_{dist}(A, \tilde{A})
\end{equation}
where $\mathcal{L}_{fair}$ measures the fairness level of the recommendations generated by $f$ with $\tilde{A}$, $\mathcal{L}_{dist}$ monitors the distance between $A$ and $\tilde{A}$, and $\beta$ is an hyperparameter to weigh the distance loss.

Fairness was explored according to the demographic parity notion on a binary setting.
Let $S$ be a recommendation utility function (e.g., NDCG~\cite{DBLP:journals/tois/JarvelinK02}), let $\mathcal{U}_1$ and $\mathcal{U}_2$ be a binary partition of $\mathcal{U}$ based on a demographic attribute, $\mathcal{L}_{fair}$ was operationalized as follows:
\begin{equation} \label{eq:lfair}
    \mathcal{L}_{fair} = \left\Vert S(\hat{R}, A^{\mathcal{U}_1}) - S(\hat{R}, A^{\mathcal{U}_2}) \right\Vert^2_2
\end{equation}
where $A^{\mathcal{U}_1}$ ($A^{\mathcal{U}_2}$) identifies the sub-adjacency matrix including the users in $\mathcal{U}_1$ ($\mathcal{U}_2$).
Fixing $S$ as the NDCG, \cite{DBLP:conf/cikm/BorattoFFMM23} adopted $\widehat{NDCG}$~\cite{10.1145/3655631,10.1145/3564285,DBLP:journals/ir/QinLL10} to approximate $S$ during the augmentation process, but used NDCG during evaluation.
Analogously, $\Delta$NDCG replaces $\mathcal{L}_{fair}$ during evaluation.
$\widehat{NDCG}$ was optimized with the items' relevance of the validation set.
$\mathcal{L}_{dist}$ was defined as $\frac{1}{2}\sigma(\sum_{i,j}|(A_{i,j} - \tilde{A}_{i,j}|^2_2)$, a distance function where $\sigma$ is a sigmoid that bounds $\mathcal{L}_{dist}$ in [0,1].

In the binary unfairness scenario, one demographic group (denoted as \emph{advantaged}) enjoys a higher recommendation utility than the other group (denoted as \emph{disadvantaged}).
The augmentation algorithm aims to improve the NDCG for the disadvantaged group (referred to as $\mathcal{U}_D$), such that it matches the NDCG for the advantaged group (referred to as $\mathcal{U}_A$).
Therefore, the augmentation pipeline targets solely the disadvantaged users, i.e. the users in $\mathcal{U}_D$.

\subsubsection{Sampling Policies} \label{subsubsec:cikm_sampling_policies}

User-item interaction graphs used in recommendation are typically sparse, where more than 90\% of potential interactions are missing from the graph.
In light of this, \cite{DBLP:conf/cikm/BorattoFFMM23} restricted $\tilde{E}$ to prevent the augmentation process from selecting edges that could potentially hinder the unfairness mitigation task.
Such scenario was enabled by five policies that sample the user and item set: \emph{Zero NDCG} (ZN) samples users of $\mathcal{U}_D$ who received recommendations with NDCG = 0 on the validation set; \emph{Low Degree} (LD) samples users of $\mathcal{U}_D$ with fewest interactions; \emph{Furthest} (FR) samples users of $\mathcal{U}_D$ who are the furthest from users of $\mathcal{U}_A$; \emph{Sparse} (SP) samples users of $\mathcal{U}_D$ mostly interacting with low-degree (niche) item nodes; \emph{Item Preference} (IP) samples items that are preferred the most by users of $\mathcal{U}_D$.
Two hyperparameters control the sample size for LD, FR, SP, IP: $\Psi_{\mathcal{U}}$ for the policies sampling from $\mathcal{U}_D$, $\Psi_{\mathcal{I}}$ for the policies sampling from $\mathcal{I}$.
For instance, LD with $\Psi_{\mathcal{U}} = 20\%$ samples 20\% of the users from $\mathcal{U}_D$ with the lowest degree.

\subsection{Experimental Setup} \label{subsec:exp-setting}

\subsubsection{Datasets}

The reproduced study only addressed the song and movie domain by respectively using Last.FM-1K~\cite{DBLP:books/daglib/0025137} and MovieLens-1M (ML1M)~\cite{DBLP:journals/tiis/HarperK16}.
Given the availability of larger datasets extracted from the Last.FM platform, we replaced Last.FM-1K with Last.FM-1M (LF1M)~\cite{DBLP:conf/ecir/BalloccuBCFM23}, a sample of Last.FM-1B~\cite{DBLP:conf/mir/Schedl16} including binary gender and age information.
We extended the experimental evaluation to the fashion domain with RentTheRunway (RENT)~\cite{DBLP:conf/recsys/MisraWM18}, and the points of interest domain with Foursquare~\cite{DBLP:journals/tist/YangZQ16,DBLP:conf/huc/YangZQC16} by considering the check-ins of New York City (FNYC) and Tokyo (FTKY) as two distinct datasets as previous works~\cite{DBLP:journals/tist/YangZQ16,DBLP:conf/huc/YangZQC16}.
FNYC and FTKY include a binary attribute to represent the gender information, while RENT a continuous variable to encode the users' age.
As in \cite{DBLP:conf/cikm/BorattoFFMM23}, age was binarized in \emph{Younger} (age $\leq$ 33) and \emph{Older} (age $>$ 33), such that Younger is over-represented like in LF1M and ML1M.
A k-core filtering step of 5 interactions for RENT, and 20 for LF1M, FNYC, and FTKY was applied.
As the reproduced work~\cite{DBLP:conf/cikm/BorattoFFMM23} and previous papers~\cite{DBLP:conf/sigir/Wang0WFC19,DBLP:conf/www/LiangKHJ18}, we adopt the temporal user-based splitting strategy~\cite{DBLP:conf/recsys/MengMMO20} with a ratio 7:1:2.
Table~\ref{tab:datasets} reports datasets statistics.

\subsubsection{Models}

In order to provide a comprehensive assessment of the reproduced approach, we considered a broader set of models employed in GCF.
Focusing on state-of-the-art systems, we discarded GCMC, being the oldest model studied in \cite{DBLP:conf/cikm/BorattoFFMM23}.
We relied on several implementations provided by Recbole~\cite{DBLP:conf/cikm/ZhaoMHLCPLLWTMF21,DBLP:conf/cikm/ZhaoHPYZLZBTSCX22}, and included recently proposed systems for GCF, namely AutoCF~\cite{DBLP:conf/www/XiaHHLYK23}, UltraGCN~\cite{DBLP:conf/cikm/MaoZXLWH21}, and SVD-GCN~\cite{DBLP:conf/cikm/PengSM22} (we consider both the parametric version SVD-GCN and the non-parametric one SVD-GCN-S).
The full list of models is listed in Table~\ref{tab:models}, where we also reported, for convenience, the venue and year each model was proposed.

The reproduced algorithm selects new edges by monitoring the fairness level as the input graph is modified.
However, if the input graph is only used during the training stage, the augmented graph does not influence the resulting recommendations during the inference stage.
Models falling in this category are labeled as non-\emph{augmentable} in Table~\ref{tab:models}.
In Section~\ref{subsec:re-training}, we re-train non-augmentable GNNs and non-GNN models with the graph augmented by means of augmentable GNNs to explore the transferability of the fair knowledge embedded in the graph.
Particularly, we select UltraGCN and SVD-GCN-S as non-augmentable GNNs, while BPR~\cite{DBLP:conf/uai/RendleFGS09}, NeuMF~\cite{DBLP:conf/www/HeLZNHC17}, ENMF~\cite{DBLP:journals/tois/ChenZZLM20}, EASE~\cite{DBLP:conf/www/Steck19}, and DiffRec~\cite{DBLP:conf/sigir/WangXFL0C23} as recommenders not based on GNNs.
They cover diverse systems families, such as matrix factorization, autoencoders, and diffusion models.

\subsubsection{Experimental Settings}

All the recommenders are trained over 100 epochs and the epoch reporting the best utility is selected according to the validation set.
An extensive grid-search strategy is used to optimize the hyper-parameters, but we follow \cite{DBLP:conf/www/0160ZZD23} in fixing some crucial parameters to conduct a fair comparison across models.
Specifically, we fix the GNNs embedding size to 64, the negative sampling to 10 items, and the message-passing depth to 3, i.e. the number of graph convolutional layers\footnote{2 graph convolution layers and 1 graph transformer layer for AutoCF} for the models based on explicit message passing~\cite{DBLP:conf/ecir/AnelliDNMPP23}.
3 layers would guarantee the models to aggregate information coming from different-type nodes~\cite{DBLP:conf/ecir/AnelliDNMPP23}.
{\color{black} Fixing these parameters guarantees a fair comparison, given that all the models encode users and items in the same latent space (for the embedding size), the models capture the same size of collaborative neighborhood (for the number of graph convolutional layers), and the models refine the users' and items' representation with the same amount of negative instances (for the negative sampling size).}

The augmentation process is optimized through the relevance judgements of the validation set and run for 800 epochs, but early stopped if $\Delta$NDCG is not reduced by $0.0001$ after 7 consecutive epochs.
In all the experiments, $\beta$ is set to $0.5$ as in \cite{DBLP:conf/cikm/BorattoFFMM23}, NDCG pertains to NDCG@$10$, i.e. measured on the $10$ items predicted as most relevant.
Following \cite{DBLP:conf/cikm/BorattoFFMM23}, we set $\Psi_{\mathcal{U}}$ to $35\%$ and $\Psi_{\mathcal{I}}$ to $20\%$.

\section{FA4GCF}

In this section, we describe the main contributions of our work that drive the provided framework, FA4GCF.
First, we formally define the sampling policies introduced in \cite{DBLP:conf/cikm/BorattoFFMM23}.
Then, we formalize the impact of the augmentation process on the models' inference step.
Finally, we present the proposed set of extended policies.

\subsection{Sampling Policy Formalization} \label{subsec:policy_formalization}

Despite their benefits, the sampling policies proposed by \cite{DBLP:conf/cikm/BorattoFFMM23} were only deduced from an empirical viewpoint.
To this end, we formally define each policy and the different scenarios of their application.

A sampling policy can be viewed as a filter $\omega$ over $\mathcal{U}_D$ or $\mathcal{I}$ based on a certain criterion.
By mutually referring to $\mathcal{U}_D$ or $\mathcal{I}$ as $Z$, we can formalize a general definition of a sampling policy:
\begin{equation}
    \bar{Z} = \{z \; | \; z \in \omega(Z, \Psi_Z)\}
\end{equation}
where $\bar{Z}$, which mutually refers to $\bar{\mathcal{U}}_D$ or $\bar{\mathcal{I}}$, is the sampled set.

Let $\text{\textbf{argmin}}_{Z'}$ be a shorthand for $\text{\textbf{argmin}}_{Z' \subset Z, |Z'|/|Z| \approx \Psi_Z}$, each policy in Section~\ref{subsubsec:cikm_sampling_policies} is formally defined by a specific filter $\omega$:
\begin{equation} \label{eq:cikm_policies}
    \begin{aligned}
        \textbf{(ZN)} \quad \omega(\mathcal{U}_D, \Psi_{\mathcal{U}}) &= \{u \; | \; u \in \mathcal{U}_D, \; \text{NDCG}(\hat{R}, A^{\{u\}}) = 0 \} \\
        \textbf{(LD)} \quad \omega(\mathcal{U}_D, \Psi_{\mathcal{U}}) &= \underset{\mathcal{U}'}{\text{\textbf{argmin}}}\sum_{u \in \mathcal{U}'}{\degree(u)} \\
        \textbf{(FR)} \quad \omega(\mathcal{U}_D, \Psi_{\mathcal{U}}) &= \underset{\mathcal{U}'}{\text{\textbf{argmax}}}\sum_{u \in \mathcal{U}'}{\sum_{u_A \in \mathcal{U}_A}{\dist(u, u_A)}} \\
        \textbf{(SP)} \quad \omega(\mathcal{U}_D, \Psi_{\mathcal{U}}) &= \underset{\mathcal{U}'}{\text{\textbf{argmin}}}\sum_{u \in \mathcal{U}'}\!\frac{1}{|\mathcal{N}(u)|} \sum_{i \in \mathcal{N}(u)} \! \degree(i) \\
        \textbf{(IP)} \hspace{11pt} \quad \omega(\mathcal{I}, \Psi_{\mathcal{I}}) &= \underset{\mathcal{I}'}{\text{\textbf{argmax}}}\frac{|\mathcal{U}|}{|\mathcal{U}_D|}\sum_{i \in \mathcal{I}'}{\frac{\sum_{u \in \mathcal{U}_D \cap \mathcal{N}(i)}{R_{u,i}}}{\sum_{u \in \mathcal{U} \cap \mathcal{N}(i)}{R_{u,i}}}} \\
    \end{aligned}
\end{equation}

where $\degree(\cdot)$ is the node degree, $\dist(\cdot, \cdot)$ is the geodesic distance between two distinct nodes, $\mathcal{N}(\cdot)$ is the neighborhood node set.

Given the above formalization, the restriction applied to the augmentation process can be described as reducing the candidate edges $\tilde{E}$ by means of the sampling policy $\omega$.
Specifically, given that the sampling policies can be applied to either the disadvantaged user set or the item set or both, three scenarios can be delineated:

\begin{equation}
\begin{tabular}{c@{\hskip 10pt}c}
     \textbf{\small User Sampling (U)} & $\tilde{E} = \{(u, i) \; | \; (u, i) \in \tilde{E}, u \in \bar{\mathcal{U}}_D \}$ \\[1pt]
     \textbf{\small Item Sampling (I)} & $\tilde{E} = \{(u, i) \; | \; (u, i) \in \tilde{E}, i \in \bar{\mathcal{I}} \}$ \\
     \textbf{\makecell{\small User+Item \\ \small Sampling} (U+I)} & $\tilde{E} = \{(u, i) \; | \; (u, i) \in \tilde{E}, u \in \bar{\mathcal{U}}_D, i \in \bar{\mathcal{I}} \}$
\end{tabular}
\end{equation}

\subsection{Augmentation Formalization} \label{subsec:aug_formalization}

We now formalize the effect of sampling policies on the augmentation algorithm.
Specifically, we will delve into the technical definition of message-passing neural networks (MPNNs)~\cite{DBLP:journals/corr/abs-2104-13478} and GNNs based on singular value decomposition (SVD-GCN)~\cite{DBLP:conf/cikm/PengSM22}, and extend them to include the augmentation operation.
GNN-based recommender systems typically optimize a latent representation $\mathbf{e} = \{\mathbf{e}_u, \mathbf{e}_i\}$~\cite{DBLP:conf/ecir/AnelliDNMPP23}, where $\mathbf{e}_u \in \mathbb{R}^{|\mathcal{U}| \times d}$ and $\mathbf{e}_i \in \mathbb{R}^{|\mathcal{I}| \times d}$ are respectively the user and item embeddings, $d$ is the embedding size.

\vspace{1mm}
\noindent\textbf{MPNNs.} The message-passing operation in MPNNs generalizes the convolution operator to irregular domains. We only show the operation for the users' representation, defined as follows:

\begin{equation} \label{eq:mpnns}
    \begin{aligned}
        \mathbf{e}_u^{(l)} &= \gamma^{(l)}\left(\mathbf{e}_u^{(l-1)}, \underset{i \in \mathcal{N}(u)}{\bigoplus} \phi^{(l)}(\mathbf{e}_u^{(l-1)}, \mathbf{e}_i^{(l-1)}) \right) \\
    \end{aligned}
\end{equation}

where $\bigoplus$ denotes a differentiable, permutation-invariant aggregation function, $\gamma$ and $\phi$ are differentiable functions.
Typically, $\bigoplus$ is realised as a nonparametric operation~\cite{DBLP:journals/corr/abs-2104-13478}, e.g., sum, mean, whereas $\gamma$ and $\phi$ are learnable.
Taking NGCF~\cite{DBLP:conf/sigir/Wang0WFC19} as an example, $\gamma$ is the LeakyReLU activation function, $\phi$ is a bi-interaction propagation layer that performs feature transformation with neighborhood and inter-dependendencies weight matrices.
As described earlier, the sampling policies generate a subset $\tilde{E}$ of candidates edges.
Then, the augmentation process in MPNNs can be written as follows:
\begin{equation} \label{eq:mpnns-aug}
    \begin{aligned}
        \mathbf{e}_u^{(l)} &= \gamma^{(l)}\left(\mathbf{e}_u^{(l-1)}, \underset{i \in \mathcal{N}(u) \cup \mathcal{N}_{\dot{E}}(u)}{\bigoplus} \phi^{(l)}(\mathbf{e}_u^{(l-1)}, \mathbf{e}_i^{(l-1)}) \right) \\
    \end{aligned}
\end{equation}

The fundamental variation is the augmented neighborhood set $\mathcal{N}_{\dot{E}}(\cdot)$, where $\dot{E} \subset \tilde{E}$ represents the edges actually added to the graph at an iteration step of the augmentation process outlined in Section~\ref{subsec:graph-aug}.
It follows that $\mathcal{N}_{\dot{E}}(\cdot)$ dynamically changes as $\dot{E}$ changes according to the update of $p$.
$\mathcal{N}_{\dot{E}}(\cdot)$ enables the message-passing operation to fine-tune the user (item) latent representation by aggregating the messages coming from unforeseen neighbors.

\vspace{1mm}
\noindent\textbf{SVD-GCN}~\cite{DBLP:conf/cikm/PengSM22}\textbf{.} This model performs an implicit message-passing operation~\cite{DBLP:conf/ecir/AnelliDNMPP23} to learn the collaborative signals from the interaction graph.
In depth, \cite{DBLP:conf/cikm/PengSM22} observes that graph convolutional networks learn a low-rank representation. 
Therefore, the $k$-largest singular values extracted by singular value decomposition (SVD) can be used to formulate the users and items latent representation as follows:
\begin{equation} \label{eq:svd-gcn}
    \begin{aligned}
        \mathbf{e}_u &= \dot{\mathrm{p}}_{k,u}^T \odot \zeta(s_k) W \\
        \mathbf{e}_i &= \dot{\mathrm{q}}_{k,i}^T \odot \zeta(s_k) W \\
    \end{aligned}
\end{equation}
where $\dot{\mathrm{p}}_{k,u}^T$ and $\dot{\mathrm{q}}_{k,i}^T$ are rows of the $k$-largest left and right singular vectors of $\dot{R}$, respectively; $s_k$ is a vector containing the $k$-largest singular values, $\odot$ is the Hadamard (element-wise) product, $\zeta$ is a function that weighs the importance of singular values.
$\dot{R}$ is the feedback matrix $R$ after a renormalization trick for solving the typical over-smoothing effect in GNNs~\cite{DBLP:conf/cikm/PengSM22}.
$\dot{R}$ is defined as follows:
\begin{equation} \label{eq:svd_r}
    \dot{R} = (D_{\mathcal{U}} + \alpha\mathrm{I})^{-\frac{1}{2}}R(D_{\mathcal{I}} + \alpha\mathrm{I})^{-\frac{1}{2}}
\end{equation}
where $\alpha \geq 0$ is a parameter introduced to shrink the gap between singular values, $D_{\mathcal{U}}$ and $D_{\mathcal{I}}$ are the diagonal degree matrices of the user and item side respectively, $\mathrm{I}$ is the identity matrix.
The augmentation process should then operate before the renormalization in \eqref{eq:svd_r} to add the predicted set $\dot{E}$ of edges.
In light of this, the augmented SVD-GCN renormalization can be formalized as:
\begin{equation} \label{eq:svd_r-aug}
    \begin{gathered}
        \dot{R} = (D_{\mathcal{U}} + \alpha\mathrm{I})^{-\frac{1}{2}}R^{\dot{E}}(D_{\mathcal{I}} + \alpha\mathrm{I})^{-\frac{1}{2}} \\
        \text{s.t.} \quad R^{\dot{E}}_{u, i} = \begin{cases}
            1  & if \; (u, i) \in \dot{E} \\
            R_{u, i}  & otherwise
        \end{cases}
    \end{gathered}
\end{equation}
In other words, $R^{\dot{E}}$ is the augmented feedback matrix that includes the edges that are actually selected for addition on the basis of $p$.

\vspace{1mm}

Therefore, the sampling policies directly impact the edges that can be iteratively added to $\dot{E}$, and, as a consequence, the available information that can be passed by neighbor nodes.
It should be noted that the post-processing nature of the augmentation algorithm devised by \cite{DBLP:conf/cikm/BorattoFFMM23} does not directly update the user/item embeddings during the iterative process.
The weights of the model $f$ are kept constant, so only the vector $p$ is learnt from the augmentation algorithm.
In other words, at each step, the predictions of $f$ are solely affected by the extended neighborhood set of each user and item.

\subsection{Extended Sampling Policies Set} \label{subsec:extended_policies}

The lack of a theoretical formalization of the sampling policies in \cite{DBLP:conf/cikm/BorattoFFMM23} is not the only weakness we examined.
Although the policies were selected by considering various factors, e.g., the interactions diversity across users, common issues addressed in the recommendation literature, and the architectures of GNNs, other aspects were overlooked.
Classical graph topological properties, e.g., the ones analyzed in \cite{DBLP:journals/corr/abs-2308-10778}, and common features typically found in user-item interaction datasets, e.g., interaction time, were not conceived as potential components that could offer valuable sampling procedures.

To address such shortcomings, we first picked the three policies that reported the most positive impact on consumer unfairness in \cite{DBLP:conf/cikm/BorattoFFMM23}, i.e. ZN, FR, IP.
Then, we devised three additional policies: two of them integrate the time dimension from a user perspective, while the other one derives from a classical graph property.

\vspace{1mm}
\noindent\textbf{Time-aware Policies.} Platforms that employ recommender systems evolve as customers perform new interactions with the items, e.g., songs, movies.
Thus, latest interaction play a key role on the model ability to recommend valuable content.
In light of this, we conceived \emph{Interaction Recency} (IR), a policy that samples users based on the recency of their latest interactions.
By restricting the augmentation process to the disadvantaged users who performed the most recent interactions, IR aims to push the advancement of the model recommendation power towards the disadvantaged users.
Following the formalization in \eqref{eq:cikm_policies}, IR can be defined as follows:
\begin{equation} \label{eq:ir}
     \omega(\mathcal{U}_D, \Psi_{\mathcal{U}}) = \underset{\mathcal{U}'}{\text{\textbf{argmax}}}\sum_{u \in \mathcal{U}'}{\max \timestamp(R_u)}
\end{equation}
where $\timestamp(\cdot)$ maps the interaction history $R_u$ of user $u$ to the time each interaction was performed.
Higher $\timestamp(\cdot)$ is more recent.

Inspired by the extensive research in consumer fairness regarding the movie~\cite{DBLP:journals/air/VassoyL24,DBLP:conf/recsys/TsintzouPT19,DBLP:conf/recsys/Steck18} and music~\cite{DBLP:journals/air/VassoyL24,DBLP:journals/ipm/MelchiorreRPBLS21,DBLP:journals/fdata/DinnissenB22} domain, we devised another sampling policy that derives from the concept of \emph{timeless} items.
This kind of items represent content that has experienced an ever-going appreciation from the customers, such as classic movies or songs that remain relevant across generations.
Therefore, we devised \emph{Item Timelessness} (IT), a policy that samples the items exhibiting the largest interval between the first and last time users interacted with them.
We deem that novel interactions with this class of items could provide more useful recommendations to the disadvantaged group.
IT is defined as follows:
\begin{equation} \label{eq:it}
     \omega(\mathcal{I}, \Psi_{\mathcal{I}}) = \underset{\mathcal{I}'}{\text{\textbf{argmax}}}\sum_{i \in \mathcal{I}'}{(\max \timestamp(R_i) - \min \timestamp(R_i))}
\end{equation}

\vspace{1mm}
\noindent\textbf{Graph Property-based Policy.} Despite the broad availability of graph topological properties, most of them were seemingly adopted in the empirical thinking illustrated in \cite{DBLP:conf/cikm/BorattoFFMM23}.
For instance, clustering coefficient~\cite{DBLP:conf/ecir/AnelliDNMPP23,DBLP:journals/socnet/LatapyMV08} might lead to similar sampling results in comparison with SP (clustering criteria are biased towards clustering together nodes with high degree~\cite{DBLP:conf/gd/Noack05}) and IP (potentially forming clusters of items preferred by distinct user segments~\cite{DBLP:journals/eswa/DiezCLB08}).
Other properties, e.g., harmonic centrality~\cite{MARCHIORI2000539}, eccentricity~\cite{DBLP:books/cu/ZM2014}, were not considered due to their similarity with other policies, e.g., FR.
To this end, we opted for \emph{pagerank}~\cite{DBLP:books/cu/ZM2014,DBLP:journals/ipl/Grolmusz15} (eigenvector centrality~\cite{DBLP:books/cu/ZM2014} was not considered due to their similarities).
Pagerank uses the random surfing assumption to recursively measure the prestige~\cite{DBLP:books/cu/ZM2014} of a node in a network, i.e. the importance of a node based on its own degree and its neighbors' degree.
Unlike other degree-based graph topological properties, e.g., degree centrality~\cite{DBLP:books/cu/ZM2014} and degree assortativity~\cite{DBLP:conf/ecir/AnelliDNMPP23,DBLP:journals/physrev/Newman03}, the inter-influence among nodes measured by pagerank is not strictly associated with the graph degree distribution.
This observation holds true as long as the graph is irregular~\cite{DBLP:journals/ipl/Grolmusz15}, i.e. the nodes have diverse degrees.
Given prevalent issues in the recommendation literature, such as the rich-get-richer effect~\cite{DBLP:conf/www/GermanoGM19,DBLP:conf/recsys/KlimashevskaiaE23} and the cold-start problem~\cite{DBLP:conf/recsys/CaoYWLPYY23,DBLP:conf/sigir/LiZLYZLCZ23}, we argue that user-item interactions networks are unlikely to be regular, and their degree distribution exhibits a scale-free behavior~\cite{DBLP:books/cu/ZM2014}.
Thus, pagerank represents a valid candidate for a sampling policy.

In our analysis, we adopt pagerank to narrow the augmentation process on the most influential nodes.
To this end, our study only envisions pagerank from an item perspective, given that the most influential items could foster information for more valuable recommendations.
Conversely, restricting the augmentation to the most influential users would benefit only them (who might already enjoy a high utility) and compromise the experience of users not labeled as relevant by pagerank.
\emph{PageRank} (PR) is defined as follows:
\begin{equation} \label{eq:pr}
     \omega(\mathcal{I}, \Psi_{\mathcal{I}}) = \underset{\mathcal{I}'}{\text{\textbf{argmax}}}\sum_{i \in \mathcal{I}'}{\pagerank_i(R)}
\end{equation}
where $\pagerank_z(\cdot)$ is the pagerank of node $z$.

\subsection{Sampling Policies Overlap}

\begin{figure}
    \centering
    \setlength{\tabcolsep}{15pt}
    \resizebox{\linewidth}{!}{
    \begin{tabular}{ccccc}
        \includegraphics{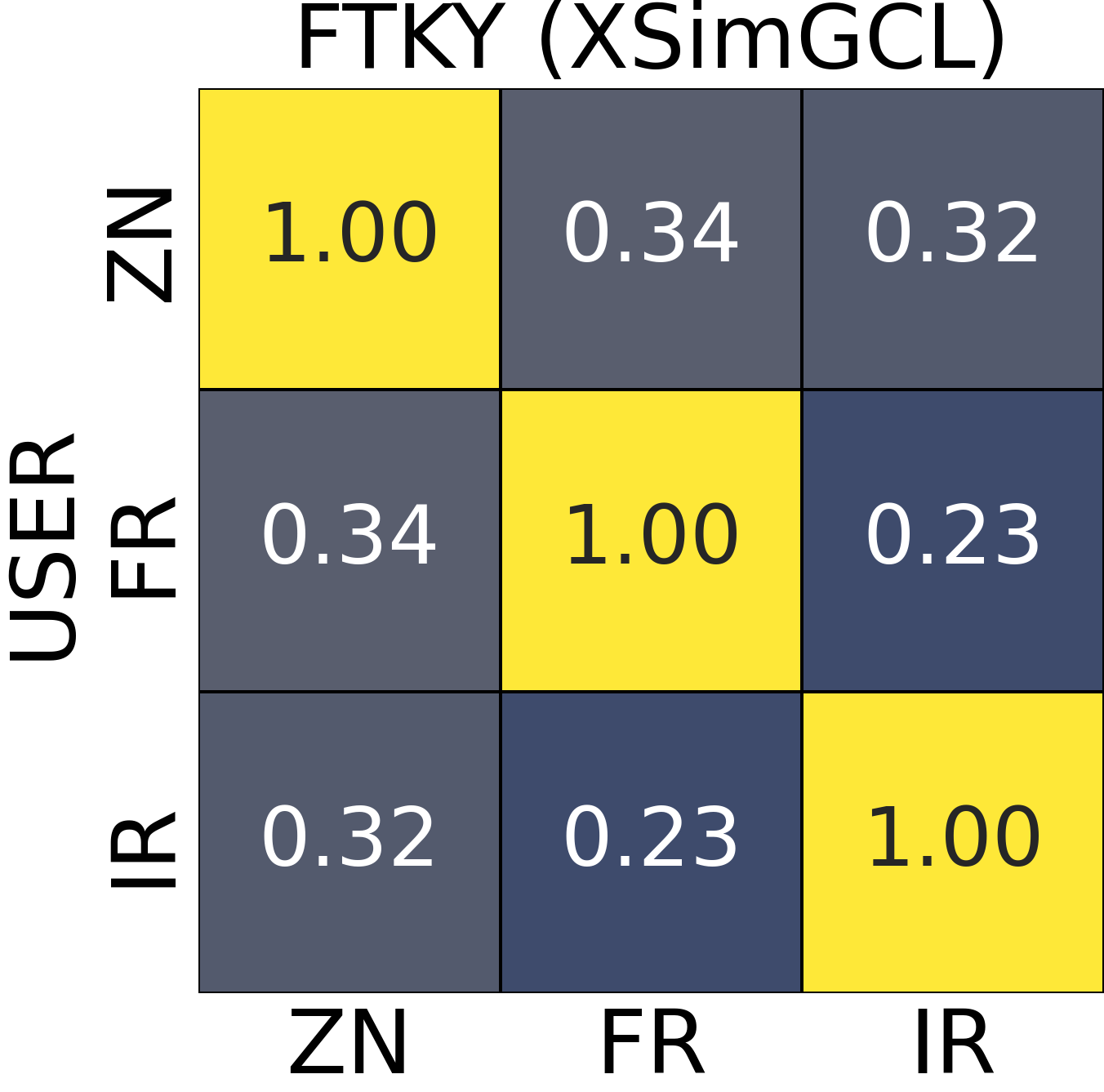} & \includegraphics{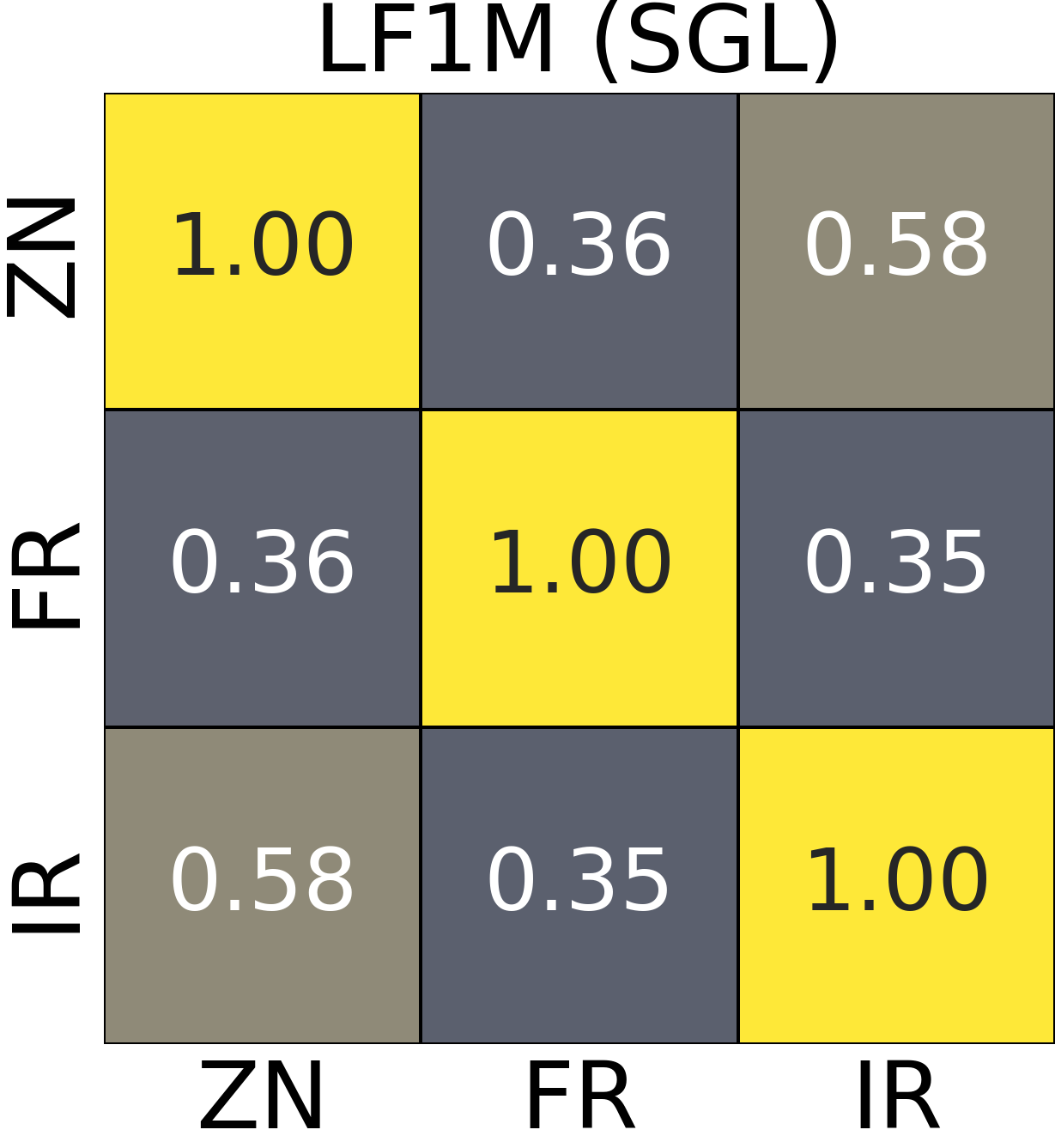} &  \includegraphics{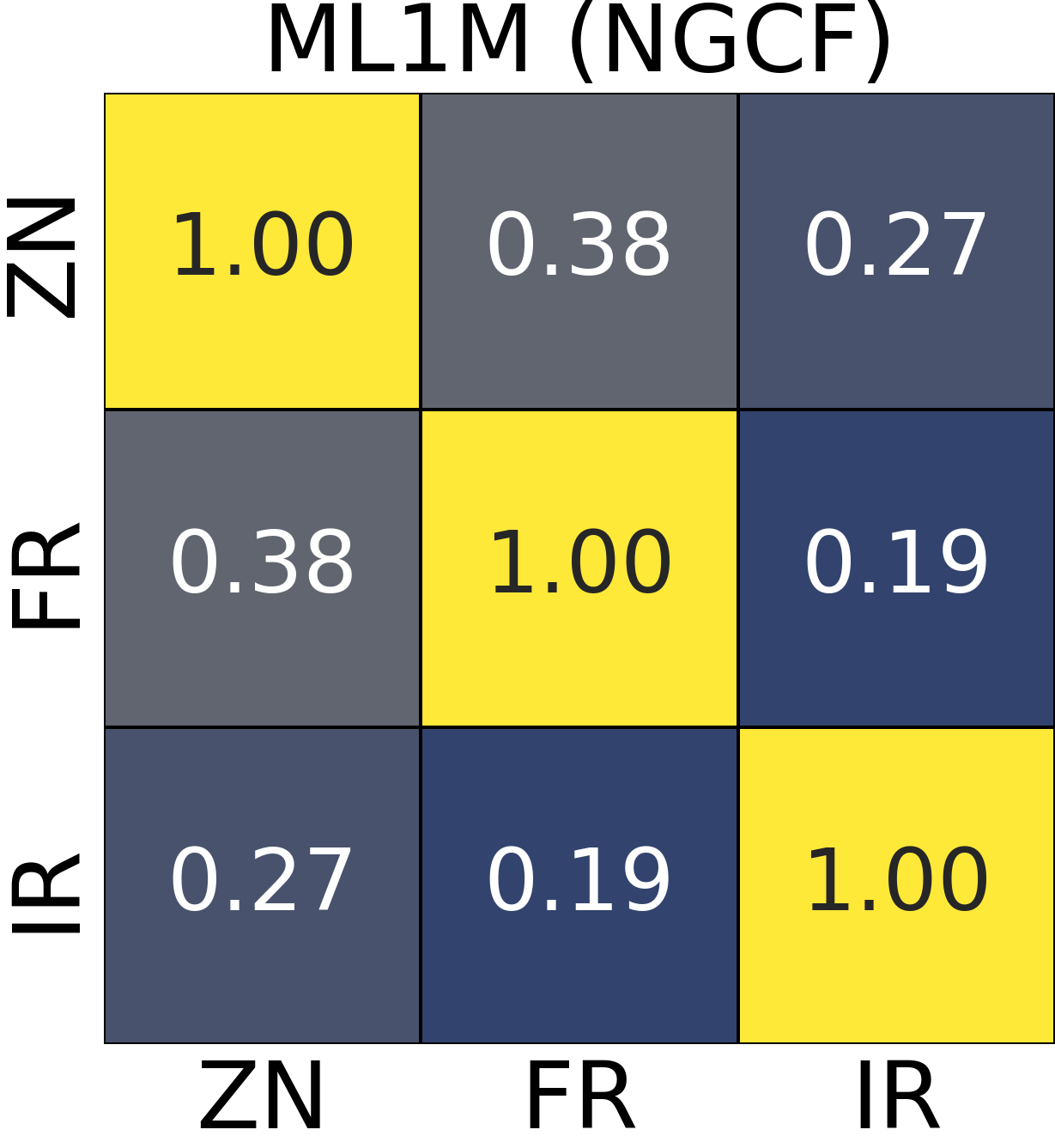} & \multirow{2}{*}[108mm]{\includegraphics{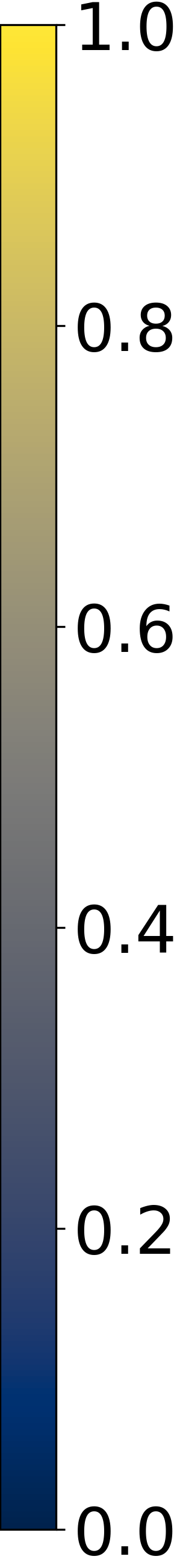}} \\[18pt]
         \includegraphics{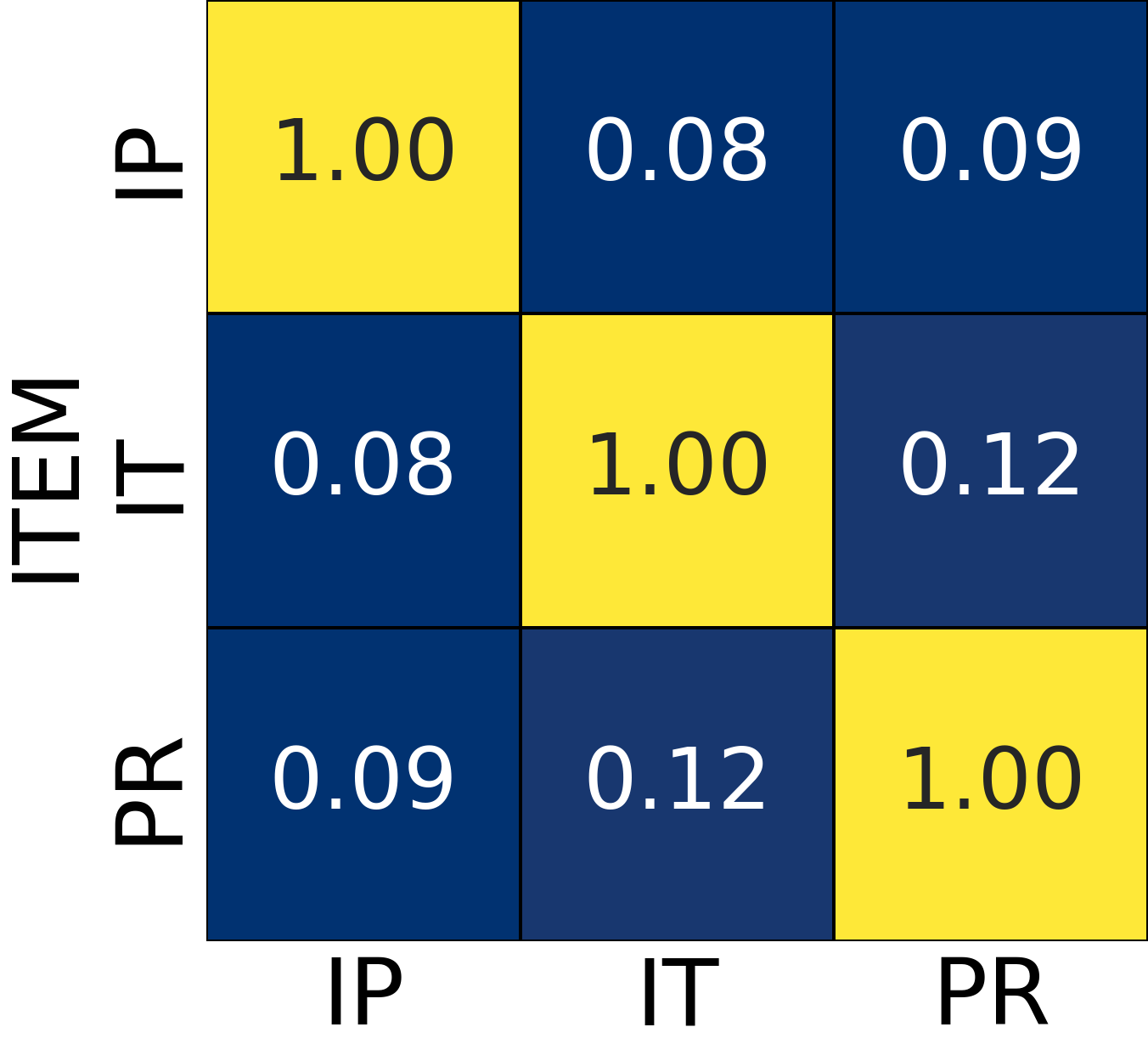} & \includegraphics{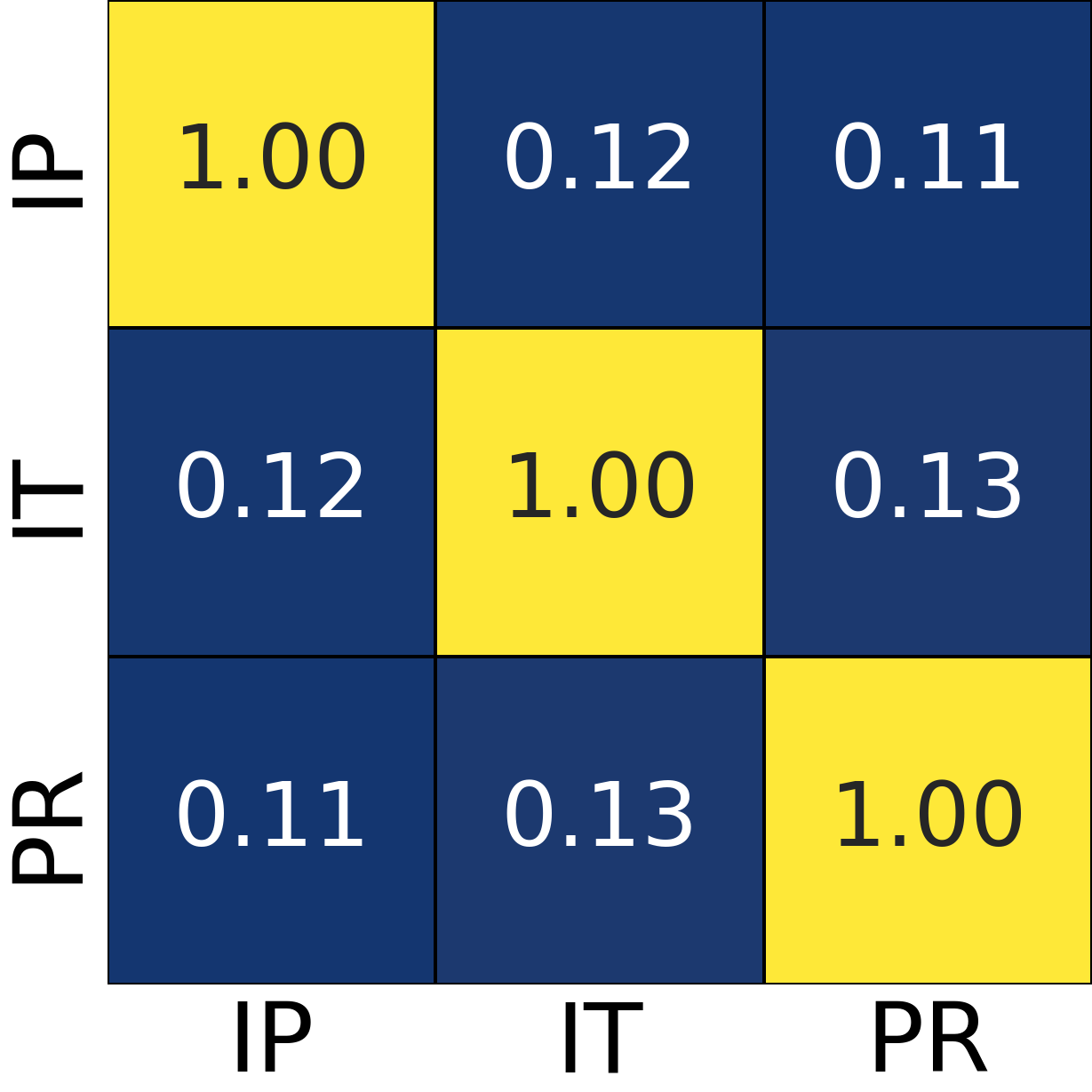} &  \includegraphics{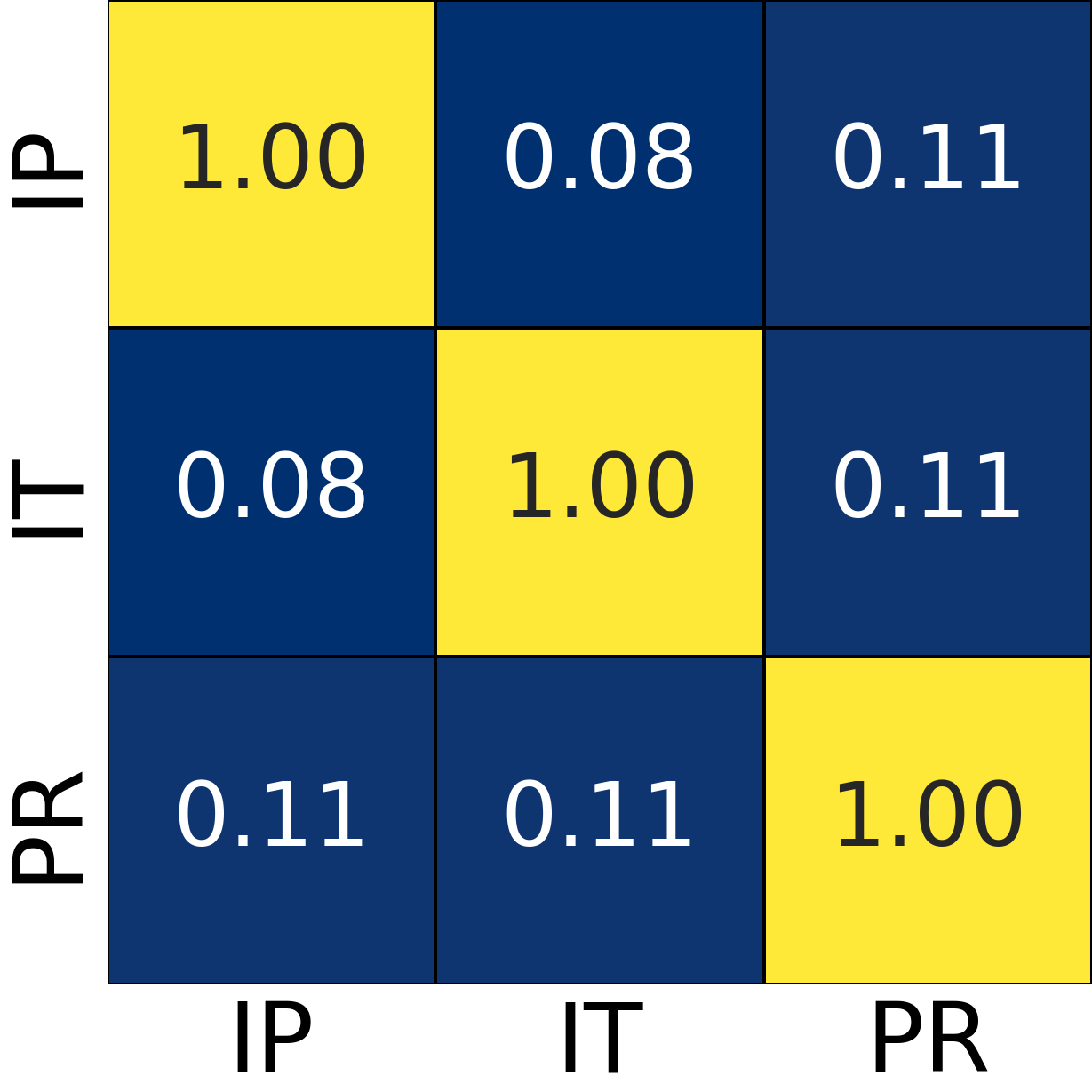} &  \\
    \end{tabular}
    }
    \caption{Jaccard similarity between sampling policies across distinct models, due to the policies being model-dependant.}
    \Description{The comparison across sampling policies in terms of Jaccard similarity, displayed as heatmaps with color comparisons, shows higher similarities for the user sampling policies, regardless of the dataset and model.}
    \label{fig:policies_overlap}
\end{figure}

\begin{table*}
    \centering
    \caption{Utility (NDCG) and fairness ($\Delta$) percentage levels before (\emph{Base}) and after (\emph{Aug}) adopting the augmented graph in the inference step of each model. The policy that generated the fairest (based on the validation set) augmented graph is reported for each setting. Underlined values denote the augmented graph led to improvements in terms of NDCG (increments) and $\Delta$ (decrements). Best NDCG and $\Delta$ values in each row are bolded. \emph{G} stands for \emph{Gender}, \emph{A} for \emph{Age}, \emph{Aug} for \emph{Augmented}.}
    \resizebox{\textwidth}{!}{
    \begin{tabular}{>{\raggedright}p{1.5mm}>{\raggedright}p{1mm}l*{9}{|>{\raggedright}p{3mm}rr}}
    \toprule
     &  &  & \multicolumn{3}{c|}{AutoCF} & \multicolumn{3}{c|}{DirectAU} & \multicolumn{3}{c|}{HMLET} & \multicolumn{3}{c|}{LightGCN} & \multicolumn{3}{c|}{NCL} & \multicolumn{3}{c|}{NGCF} & \multicolumn{3}{c|}{SGL} & \multicolumn{3}{c|}{SVD-GCN} & \multicolumn{3}{c}{XSimGCL} \\
     &  &  &  & NDCG $\uparrow$ & $\Delta$ $\downarrow_0$ &  & NDCG $\uparrow$ & $\Delta$ $\downarrow_0$ &  & NDCG $\uparrow$ & $\Delta$ $\downarrow_0$ &  & NDCG $\uparrow$ & $\Delta$ $\downarrow_0$ &  & NDCG $\uparrow$ & $\Delta$ $\downarrow_0$ &  & NDCG $\uparrow$ & $\Delta$ $\downarrow_0$ &  & NDCG $\uparrow$ & $\Delta$ $\downarrow_0$ &  & NDCG $\uparrow$ & $\Delta$ $\downarrow_0$ &  & NDCG $\uparrow$ & $\Delta$ $\downarrow_0$ \\ 
    \midrule
    & & & & & & & & & & & & & & & & & & & & & & & & & & & & \\[-1em]
    \multirow[c]{2}{*}{\rotatebox[origin=c]{90}{FNYC}} & \multirow[c]{2}{*}{G} & Base &  & 5.85 & 0.59 &  & 5.94 & 1.09 &  & 7.76 & \textbf{0.01} &  & 7.80 & 0.10 &  & 7.93 & 0.28 &  & 7.88 & 0.24 &  & 8.04 & 0.09 &  & 5.52 & 0.38 &  & \textbf{8.29} & 0.25 \\
     &  & Aug & {\small \emph{ZN+IT}} & 5.46 & $^*$\underline{\textbf{0.07}} & {\small \emph{ZN+IP}} & 5.94 & 1.12 & {\small \emph{IR+IT}} & \underline{7.89} & 0.74 & {\small \emph{ZN+PR}} & \underline{7.91} & $^*$0.23 & {\small \emph{IR+IP}} & \underline{8.13} & $^*$0.33 & {\small \emph{FR+IT}} & $^*$\underline{8.39} & $^*$1.30 & {\small \emph{IR+IT}} & \underline{8.22} & 0.60 & {\small \emph{IR+PR}} & $^*$4.26 & $^*$0.42 & {\small \emph{IR+IP}} & \underline{\textbf{8.76}} & $^*$1.22 \\
     & & & & & & & & & & & & & & & & & & & & & & & & & & & & \\[-1em]
    \cline{1-30} \cline{2-30}
    & & & & & & & & & & & & & & & & & & & & & & & & & & & & \\[-1em]
    \multirow[c]{2}{*}{\rotatebox[origin=c]{90}{FTKY}} & \multirow[c]{2}{*}{G} & Base &  & 7.05 & 1.25 &  & 7.47 & 0.51 &  & 8.85 & 0.25 &  & 8.69 & 0.31 &  & 9.10 & 0.29 &  & 9.24 & 0.36 &  & 9.41 & \textbf{0.15} &  & 9.14 & 0.23 &  & \textbf{9.82} & 0.17 \\
     &  & Aug & {\small \emph{ZN+IP}} & \underline{7.22} & 1.43 & {\small \emph{ZN+IT}} & 7.47 & 0.51 & {\small \emph{IR+IP}} & \underline{9.12} & $^*$\underline{0.24} & {\small \emph{FR}} & 8.61 & $^*$\underline{\textbf{0.08}} & {\small \emph{ZN+IP}} & 9.06 & $^*$0.31 & {\small \emph{IR+IT}} & \underline{9.39} & $^*$0.55 & {\small \emph{FR}} & \underline{10.11} & 1.38 & {\small \emph{IR+PR}} & $^*$5.40 & $^*$0.95 & {\small \emph{FR+PR}} & $^*$\underline{\textbf{10.26}} & $^*$0.88 \\
     & & & & & & & & & & & & & & & & & & & & & & & & & & & & \\[-1em]
    \cline{1-30} \cline{2-30}
    \multirow[c]{4}{*}{\rotatebox[origin=c]{90}{LF1M}} & \multirow[c]{2}{*}{A} & Base &  & 19.43 & 3.16 &  & 17.09 & \textbf{2.19} &  & 18.32 & 3.04 &  & 17.27 & 2.28 &  & 14.88 & 2.73 &  & 17.98 & 2.89 &  & 19.82 & 3.04 &  & 19.07 & 3.00 &  & \textbf{21.14} & 2.99 \\
     &  & Aug & {\small \emph{IR}} & \underline{19.81} & \underline{2.02} & {\small \emph{ZN+PR}} & 17.09 & \underline{2.17} & {\small \emph{ZN}} & 18.12 & $^*$\underline{1.95} & {\small \emph{IR+PR}} & \underline{17.42} & \underline{\textbf{1.41}} & {\small \emph{FR+IT}} & \underline{15.06} & \underline{1.69} & {\small \emph{IT}} & 17.98 & 2.89 & {\small \emph{IP}} & \underline{20.14} & $^*$\underline{2.13} & {\small \emph{IR+IP}} & $^*$14.99 & $^*$\underline{2.53} & {\small \emph{IT}} & \underline{\textbf{21.51}} & \underline{1.65} \\
    \cline{2-30}
     & \multirow[c]{2}{*}{G} & Base &  & 19.43 & 3.31 &  & 17.09 & 3.02 &  & 18.32 & 3.09 &  & 17.27 & 2.95 &  & 14.88 & \textbf{1.77} &  & 17.98 & 3.38 &  & 19.82 & 2.77 &  & 19.07 & 2.93 &  & \textbf{21.14} & 3.17 \\
     &  & Aug & {\small \emph{IR}} & \underline{19.96} & $^*$\underline{0.63} & {\small \emph{FR}} & 17.09 & 3.02 & {\small \emph{ZN+IP}} & 18.15 & $^*$\underline{1.18} & {\small \emph{FR+IP}} & 17.14 & $^*$\underline{1.39} & {\small \emph{FR+PR}} & \underline{15.01} & $^*$\underline{\textbf{0.45}} & {\small \emph{FR+IT}} & 17.98 & 3.39 & {\small \emph{PR}} & 19.81 & 2.78 & {\small \emph{FR+PR}} & $^*$15.35 & $^*$\underline{1.75} & {\small \emph{ZN+PR}} & \underline{\textbf{21.47}} & \underline{1.03} \\
    \cline{1-30} \cline{2-30}
    \multirow[c]{4}{*}{\rotatebox[origin=c]{90}{ML1M}} & \multirow[c]{2}{*}{A} & Base &  & 7.14 & 1.15 &  & 10.24 & \textbf{0.82} &  & 12.19 & 2.01 &  & 12.54 & 2.07 &  & 12.36 & 2.39 &  & \textbf{13.44} & 2.57 &  & 12.51 & 1.95 &  & 13.20 & 2.75 &  & 12.03 & 1.93 \\
     &  & Aug & {\small \emph{FR}} & $^*$6.57 & $^*$\underline{\textbf{0.63}} & {\small \emph{ZN+IP}} & 10.24 & 0.82 & {\small \emph{IR}} & \underline{12.60} & \underline{0.82} & {\small \emph{FR}} & \underline{13.06} & \underline{0.92} & {\small \emph{IR+PR}} & \underline{12.85} & \underline{1.11} & {\small \emph{ZN}} & \underline{\textbf{13.69}} & $^*$\underline{1.96} & {\small \emph{PR}} & \underline{12.81} & \underline{1.32} & {\small \emph{ZN+PR}} & $^*$6.07 & $^*$\underline{0.82} & {\small \emph{ZN}} & 12.02 & 1.94 \\
    \cline{2-30}
     & \multirow[c]{2}{*}{G} & Base &  & 7.14 & \textbf{0.77} &  & 10.24 & 1.27 &  & 12.19 & 1.49 &  & 12.54 & 1.70 &  & 12.36 & 1.93 &  & \textbf{13.44} & 2.27 &  & 12.51 & 1.78 &  & 13.20 & 2.20 &  & 12.03 & 1.71 \\
     &  & Aug & {\small \emph{ZN+PR}} & $^*$0.04 & $^*$\underline{\textbf{0.01}} & {\small \emph{ZN+PR}} & 10.24 & 1.27 & {\small \emph{FR+PR}} & \underline{12.65} & \underline{0.16} & {\small \emph{FR+PR}} & \underline{13.00} & $^*$\underline{0.06} & {\small \emph{ZN+IT}} & \underline{12.53} & \underline{1.23} & {\small \emph{ZN+IP}} & \underline{\textbf{13.72}} & $^*$\underline{1.25} & {\small \emph{IR+IP}} & \underline{13.01} & \underline{0.03} & {\small \emph{ZN+IT}} & $^*$6.13 & $^*$\underline{0.41} & {\small \emph{ZN+IT}} & \underline{12.16} & \underline{1.13} \\
    \cline{1-30} \cline{2-30}
    \multirow[c]{2}{*}{\rotatebox[origin=c]{90}{RENT}} & \multirow[c]{2}{*}{A} & Base &  & 0.91 & 0.36 &  & 0.71 & 0.48 &  & 0.79 & 0.29 &  & 0.81 & 0.26 &  & 0.86 & 0.33 &  & 0.75 & \textbf{0.05} &  & 1.01 & 0.40 &  & 0.98 & 0.26 &  & \textbf{1.03} & 0.67 \\
     &  & Aug & {\small \emph{IR+IP}} & \underline{0.94} & 0.44 & {\small \emph{IR+IP}} & 0.71 & 0.48 & {\small \emph{IR+IT}} & 0.79 & 0.29 & {\small \emph{IP}} & 0.81 & 0.26 & {\small \emph{IT}} & \underline{0.87} & 0.34 & {\small \emph{FR+PR}} & \underline{0.78} & 0.15 & {\small \emph{ZN+IP}} & \underline{1.02} & 0.42 & {\small \emph{IT}} & 0.89 & \underline{\textbf{0.14}} & {\small \emph{ZN+PR}} & \underline{\textbf{1.04}} & 0.71 \\
     & & \multicolumn{1}{r}{} & & & \multicolumn{1}{r}{} & & & \multicolumn{1}{r}{} & & & \multicolumn{1}{r}{} & & & \multicolumn{1}{r}{} & & & \multicolumn{1}{r}{} & & & \multicolumn{1}{r}{} & & & \multicolumn{1}{r}{} & & & \multicolumn{1}{r}{} & & & \\[-1em]
    \bottomrule
    \addlinespace[3pt]
    \multicolumn{30}{l}{\large $^*$Statistically different from the \emph{Base} value under a Wilcoxon signed-rank test with a 95\% confidence interval.}
    \end{tabular}
    }
    \label{tab:overall_performance}
    \vspace{-2mm}
\end{table*}

Before going through the experiments on the reproduced work, we conduct a crucial analysis that was not included in \cite{DBLP:conf/cikm/BorattoFFMM23}.
Specifically, we assess whether two policies of the same type overlap, i.e. they actually sample subsets of users or items as disjoint as possible.
Such a scenario guarantees that applying different sampling policies would result in augmentation procedures that address different portions of the graph.
Figure~\ref{fig:policies_overlap} reports the Jaccard similarity across sampling policies for each policy type on the largest datasets.
For the user policy type, we focus only on gender, since it is included in most of the datasets.
Due to the ZN sampling being non-deterministic and to the labeling of a demographic group as advantaged being dependent on the model, the Jaccard similarity for each dataset is taken from distinct models, prioritizing the ones reporting high utility levels (see Table~\ref{tab:overall_performance}).

Figure~\ref{fig:policies_overlap} clearly shows a higher overlap across user-sampling policies than item-sampling ones.
This is mainly attributed to the value of $\Psi_{\mathcal{U}}$ being greater than $\Psi_{\mathcal{I}}$, which results in a more likely overlap.
Nonetheless, the maximum similarity across user-sampling policies guarantees that at least 60\% of users are distinct for each augmentation process in most of the settings, except for ZN and IR under LF1M.
The high overlap between these two policies underscores that users with the most recent interactions in LF1M (IR) are not satisfied by the system (ZN).
High similarities are also generally reported between ZN and FR.
It follows that a significant fraction of the disadvantaged users that are the furthest from the advantaged ones (FR) are receiving low-utility recommendations (ZN).

Overall, the similarities in Figure~\ref{fig:policies_overlap} prove the selected policies will lead to augmentation trials targeting diverse graph portions and resulting in distinct effects on utility and fairness.

\section{Experimental Evaluation}

In this section, we evaluate the unfairness mitigation effectiveness of the reproduced method by conducting a suite of experiments aimed to answer the following research questions:
\begin{enumerate}[start=1, label={\bfseries RQ\arabic*:}]
    \item To what extent the augmented graph impacts the utility and fairness levels of state-of-the-art models in GCF?
    \item How the level of unfairness mitigation varies across sampling policies? 
    \item What is the impact of sampling a smaller or larger set of users/items for the augmentation process?
    \item Can the fair knowledge embedded in the augmented dataset be transferred to other models during re-training?
\end{enumerate}

\subsection{RQ1: Utility-Fairness Benchmark}

As a first experiment, we aim to discover if the reproduced mitigation procedure can effectively counteract consumer unfairness, and result in fair and high-utility recommendations.
Hence, we test the algorithm on all augmentable GNNs and analyze the impact of the augmented graph on utility and fairness over all datasets.
Table~\ref{tab:overall_performance} reports the utility (NDCG) and fairness ($\Delta$, shorthand for $\Delta$NDCG) levels of the recommendations generated with the original (\emph{Base}) and augmented graph (\emph{Aug}).
The values labeled as \emph{Aug} pertain to the policies that reported the lowest $\Delta$ on the validation set.

A notable result is reported on SGL under ML1M, where not only the unfairness was optimally mitigated, but also the overall NDCG was increased.
Such a result suggests the augmentation process successfully increased the utility of the disadvantaged group to match that of the advantaged group.
The consistency reported under LF1M and ML1M on HMLET, LightGCN, NCL, SGL (except for LF1M on gender), and XSimGCL (except for ML1M on age) underlines the augmentation process can specifically favor high-performing models, leading to fairer and more useful recommendations.
SVD-GCN does not follow the same trend as the other models, despite exhibiting high utility.
This might be caused by the frozen weight being specifically optimized for the singular values of the \emph{Base} graph, which negatively affects the contribution of the singular values extracted from the decomposition of the augmented graph.

The least-performing models, AutoCF and DirectAU, follow trends different from the other GNNs, resulting in improvements in either NDCG or $\Delta$ or no minimal effect.
AutoCF is positively affected by the augmented graph in some settings, but the \emph{Base} model achieved state-of-the-art utility only under LF1M and RENT.
Particularly under LF1M, the augmentation process effectively mitigated the unfairness and increased the overall NDCG.
The alignment and uniformity constraints adopted by DirectAU cause the augmentation algorithm to select just a few edges before being interrupted by the early stopping criterion.
The few added edges minimally affect the resulting recommendations, exhibiting DirectAU as an inappropriate choice for the reproduced mitigation procedure.

The highlighted observations suggest that larger datasets benefit from the augmented graph to a greater extent than smaller ones.
This can be attributed to the capacity of the mitigation procedure to learn the unfairness trend from a larger amount of data.
This data size factor also contributes to the consistency of the unfairness trend across the validation and test set, whereas smaller datasets might present a different unfairness distribution across the two sets.
{\color{black} While these aspects fully address the experiments under RENT, we argue that other data nuances might have also influenced the augmentation algorithm, especially considering that the smaller Foursquare datasets (FNYC and FTKY) share the same minimum user degree as the larger LF1M and ML1M datasets.
Given that the augmentation algorithm aims to increase the utility of the recommendations for a subgroup of users (the disadvantaged group), we argue that our method's effectiveness significantly depends on the presence of a strongly clustered structure within the targeted subgraph.
A clustered subgraph enables the additional edges to specifically target the disadvantaged group, increasing the utility of their recommendations while minimally affecting the advantaged group.
Therefore, it is reasonable to consider that the Foursquare datasets might not exhibit the described structure, as venue check-ins might not form common preference clusters like those towards movie and song genres, as in ML1M and LF1M, respectively.
}

The table also emphasizes that the NDCG scoreboard is dominated by just 2 models out of 9.
Specifically, NGCF dominates the scene on ML1M, XSimGCL on FNYC, FTKY, LF1M, and RENT.
Still, it is evident that GNN-based recommenders by now consolidated as baselines~\cite{DBLP:conf/recsys/AnelliMPBSN23}, such as LightGCN and NGCF, exhibit comparable utility with respect to state-of-the-art models, e.g., XSimGCL.

Overall, the reproduced approach represents a powerful solution to both mitigate unfairness and increase utility for the GCF landscape, especially under datasets with a large amount of interactions.

\subsection{RQ2: Comparison across Policies} \label{sec:comparison_across_policies}

The previous experiment proved that the reproduced method can effectively generalize across different models and large datasets.
Nonetheless, such results did not highlight which sampling policies better captured the information that was missing from the original graph to result in fairer recommendations.
To this end, we compare the mitigation level of each sampling policy under the same setting.
Given that the mitigation procedure successfully mitigated unfairness on HMLET and LightGCN under almost all datasets, we restrict the following assessment to such models.
Additionally, we only consider unfairness across groups defined by the gender attribute, which is included in most of the examined datasets.

\begin{figure}
    \centering
    \setlength{\tabcolsep}{15pt}
    \resizebox{1.02\linewidth}{!}{
    \fontsize{32pt}{32pt}\selectfont
    \begin{tabular}{p{0.2cm}ccccc}
        & \hspace{5mm} FNYC & \hspace{5mm} FTKY & \hspace{5mm} LF1M & \hspace{5mm} ML1M & \\[0.1pt]
        \multirow{1}{*}[55mm]{\rotatebox[origin=c]{90}{HMLET}} & \includegraphics{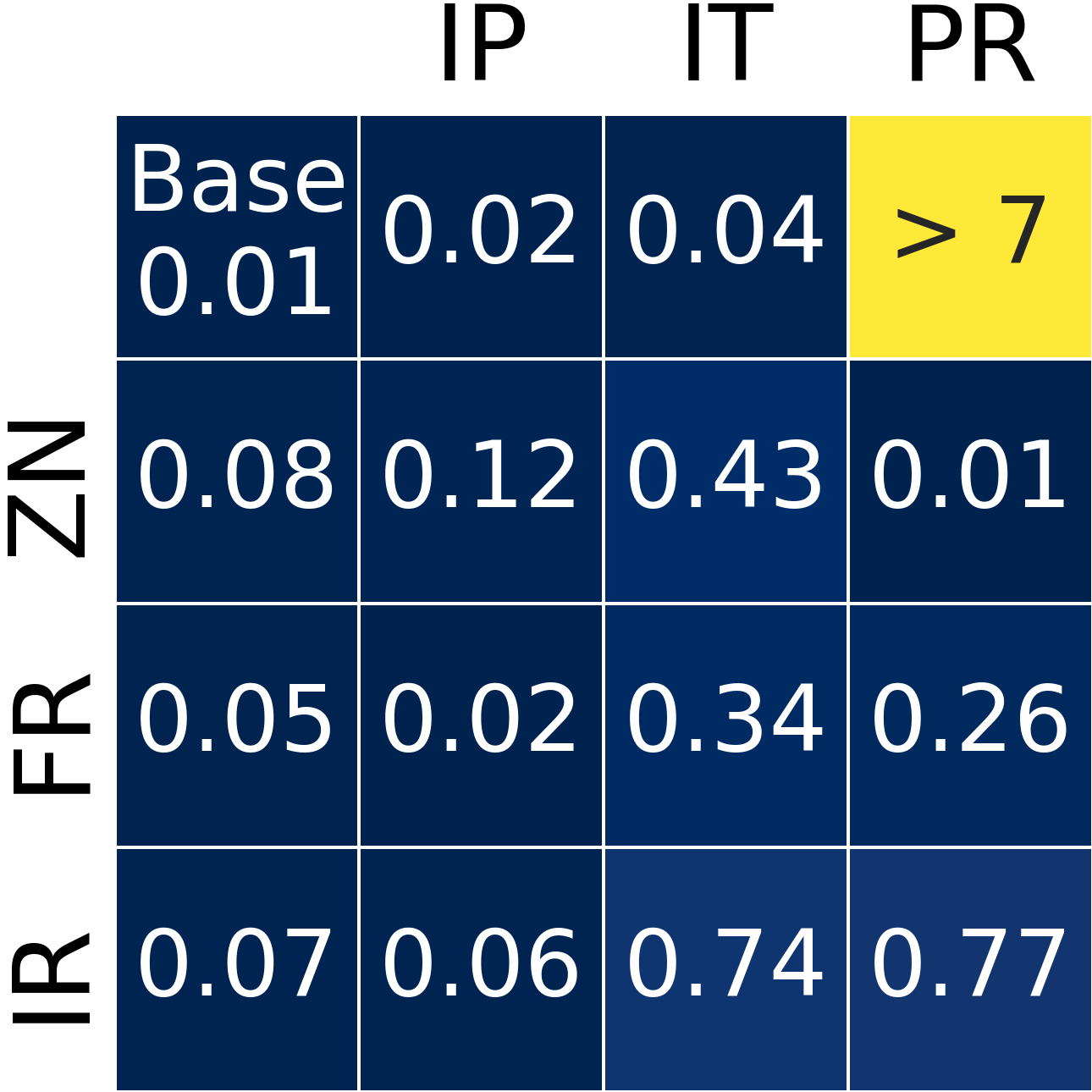} & \includegraphics{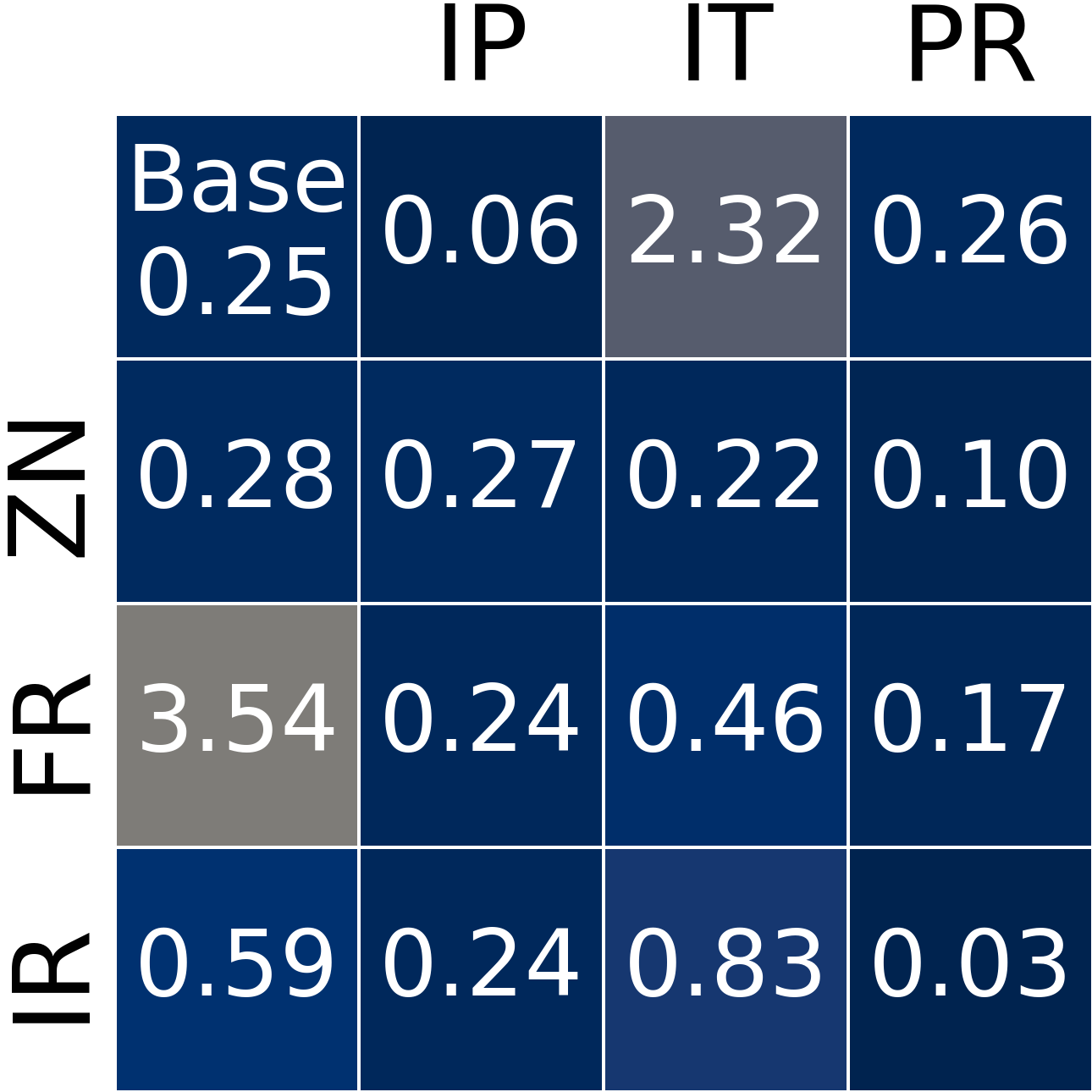} & \includegraphics{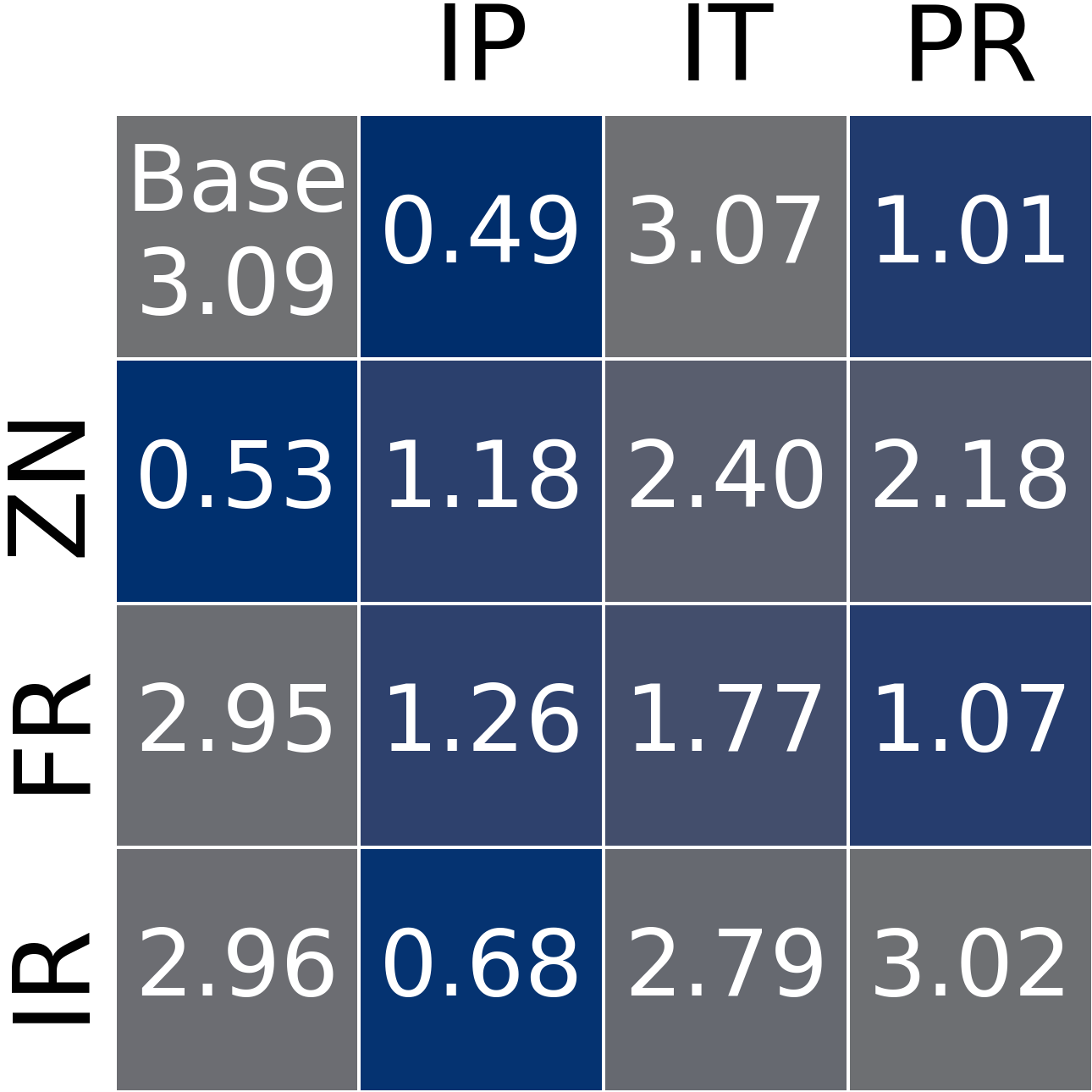} &  \includegraphics{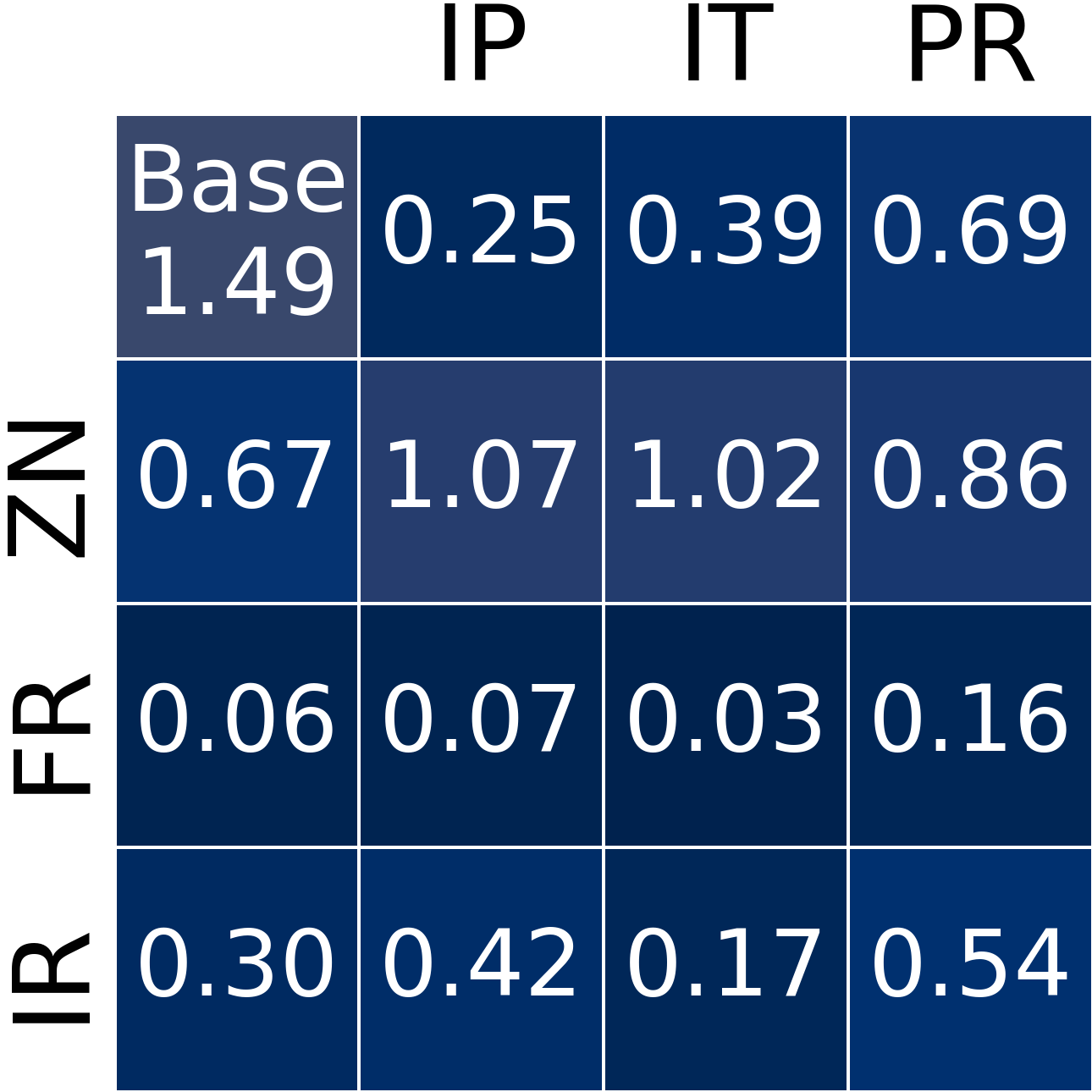} & \multirow{2}{*}[92mm]{\includegraphics{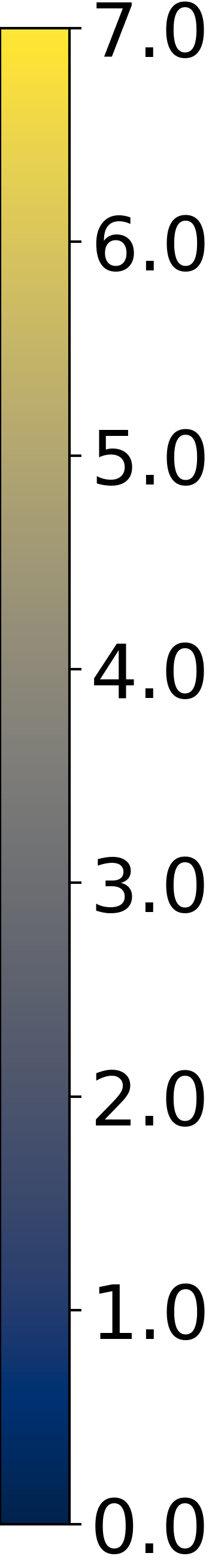}} \\[16pt]
        \multirow{1}{*}[62mm]{\rotatebox[origin=c]{90}{LightGCN}} & \includegraphics{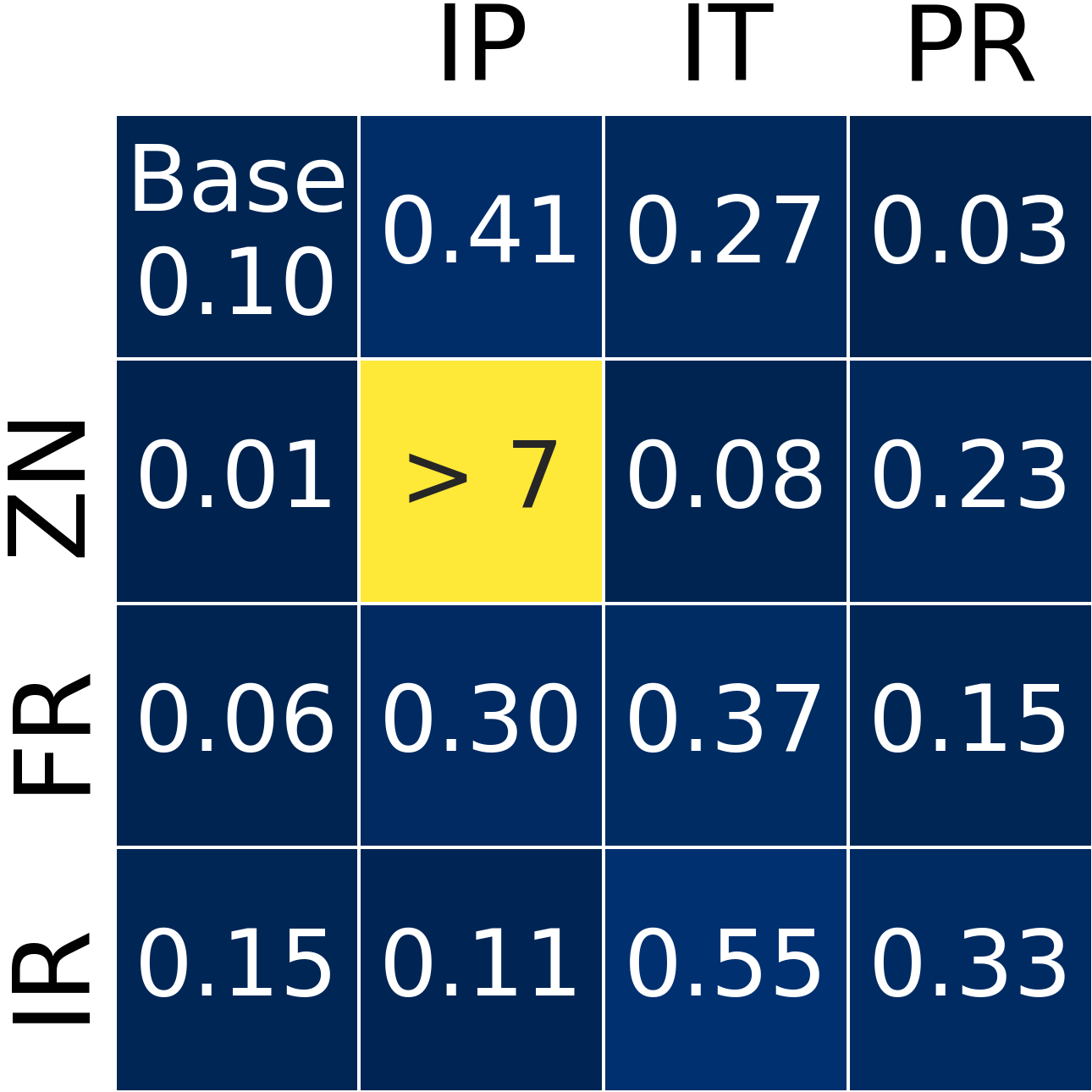} & \includegraphics{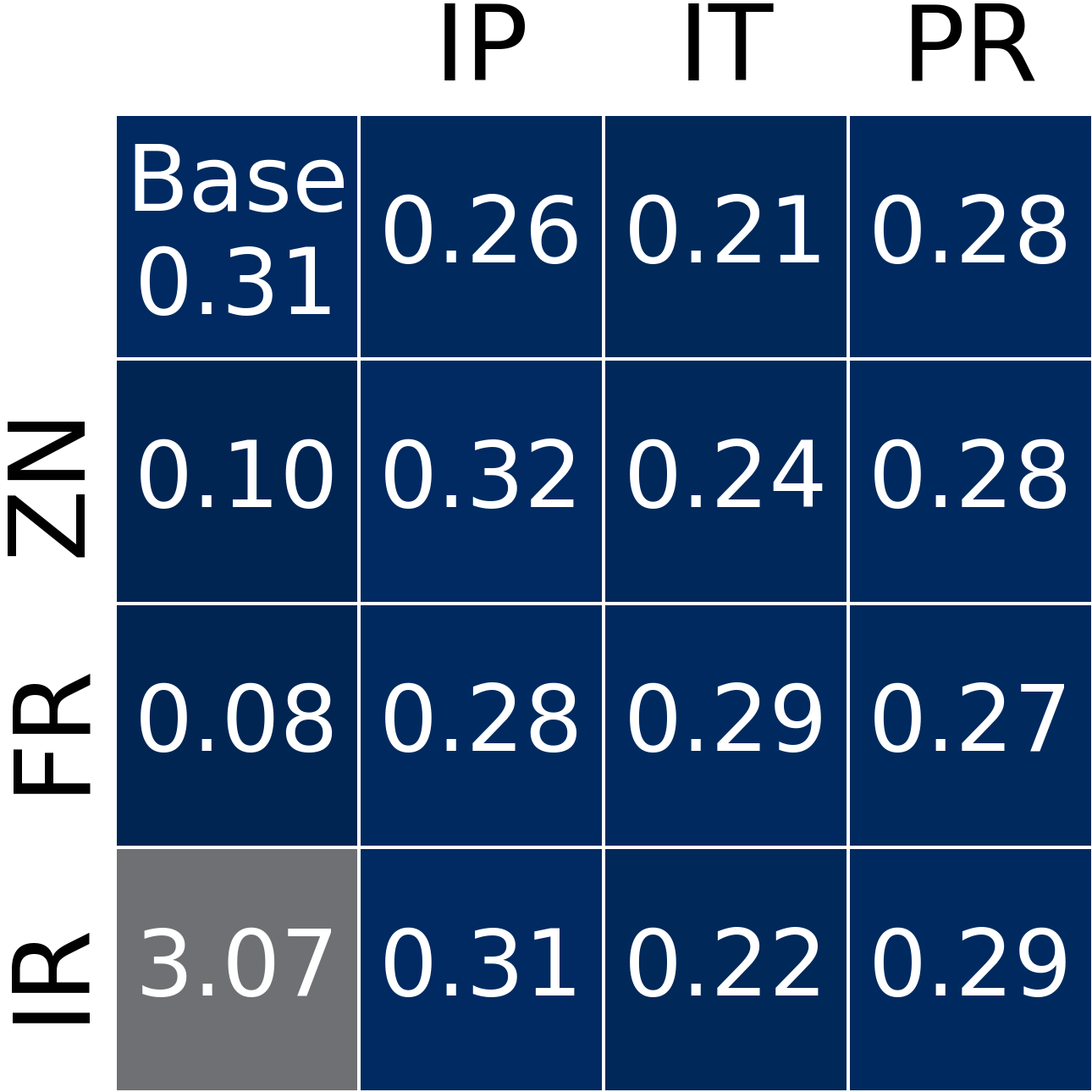} & \includegraphics{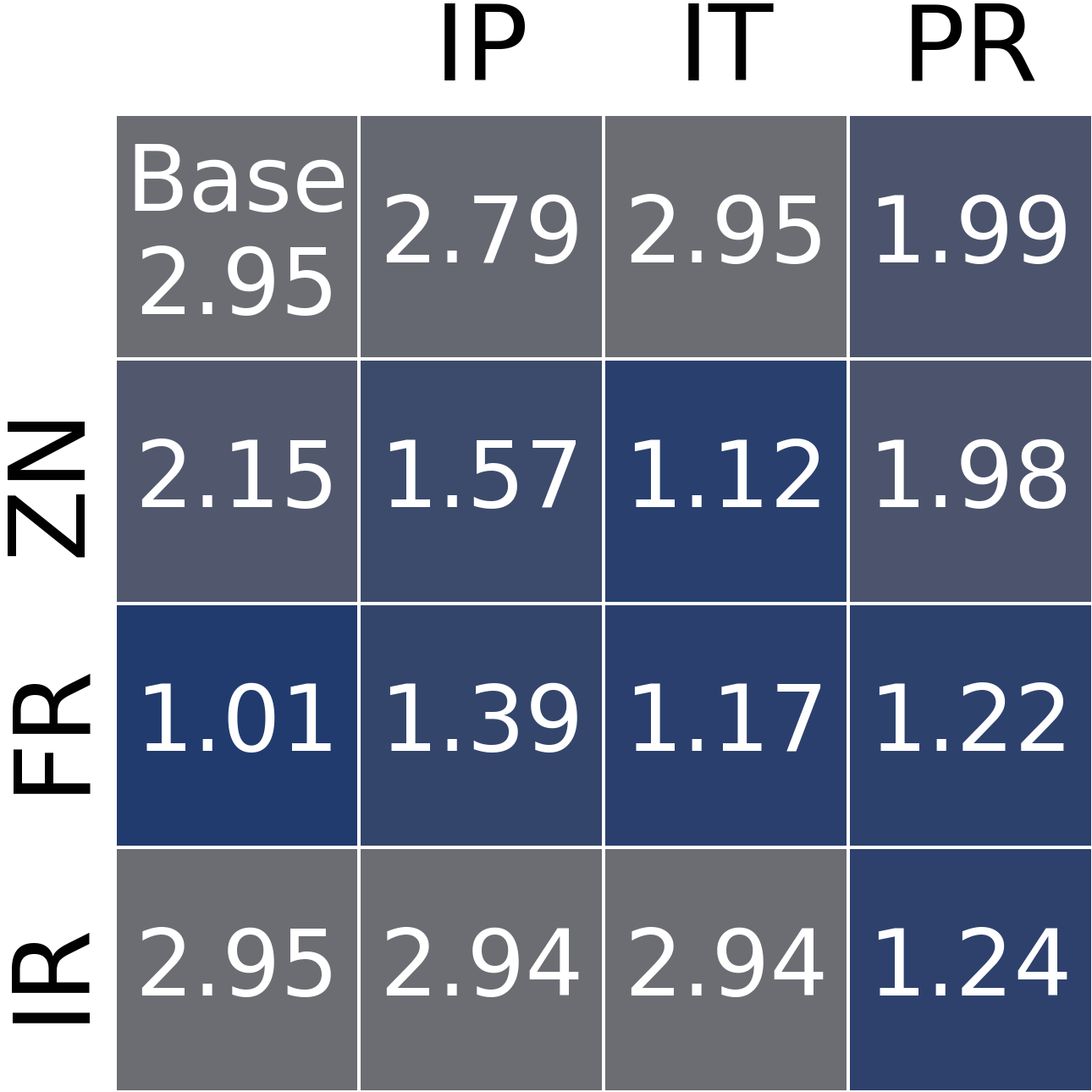} &  \includegraphics{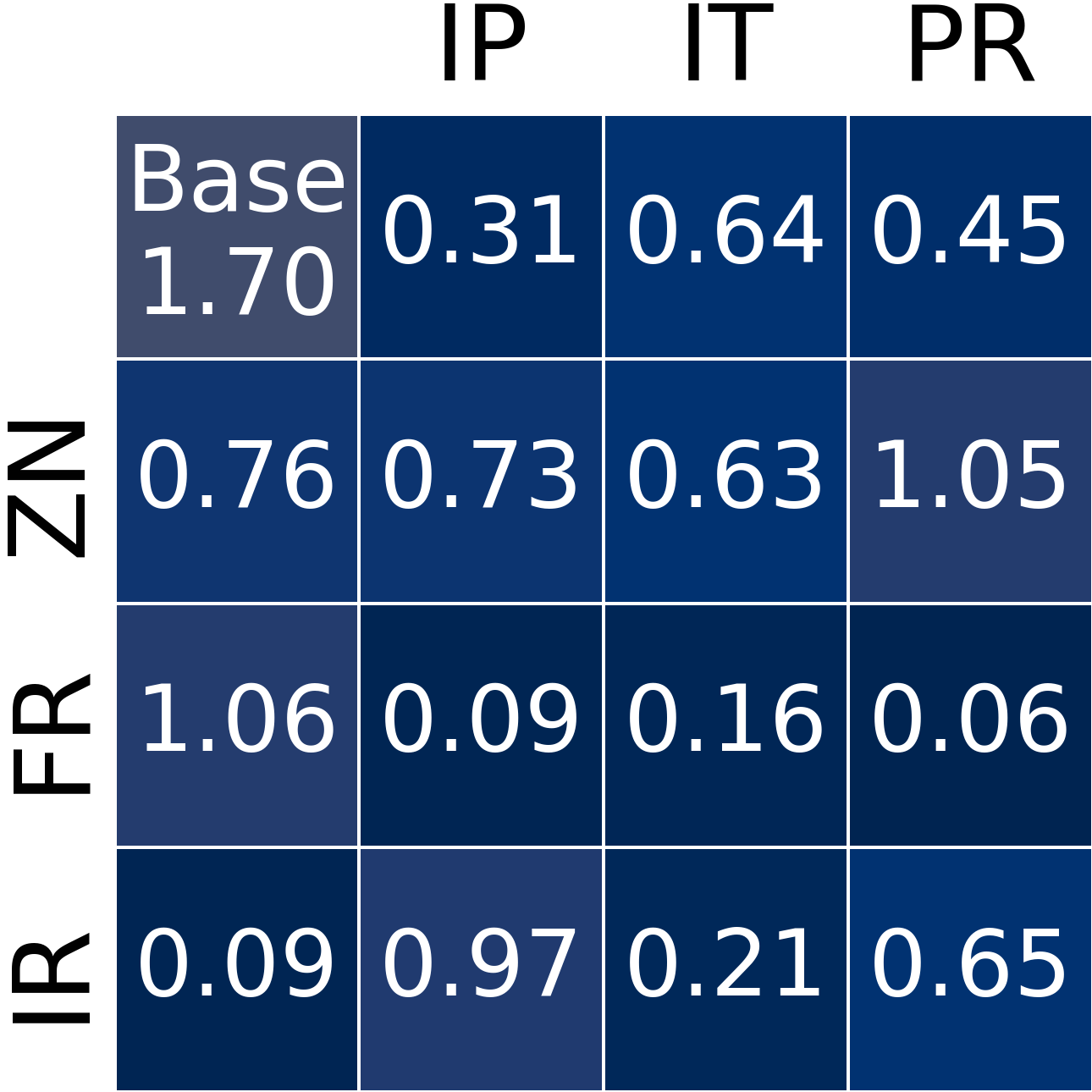} &  \\
    \end{tabular}
    }
    \caption{Comparison among sampling policies in terms of unfairness mitigation ($\Delta$ percentage) across gender groups. The $\Delta$ of the base model is labeled as \emph{Base}. The first column (row) of each subplot pertains to the User-sampling (Item-sampling) settings. The other cells pertain to the User-Item sampling settings. Smaller values, i.e. darker cells, are better.}
    \Description{Heatmaps that display the comparison across sampling policies in terms of unfairness mitigation, where fairer levels are represented by cells with lighter colors compared with the Base one at the top-left of each heatmap.}
    \label{fig:policies_comparison}
\end{figure}

Figure~\ref{fig:policies_comparison} reports the $\Delta$ resulting from the augmentation process applied with each policy setting and the base model (cell labeled as \emph{Base}).
As previously highlighted, larger datasets benefit more from the reproduced mitigation procedure compared with smaller corpora.
The subplots in Figure~\ref{fig:policies_comparison} emphasize this finding, given that $\Delta$ reductions are mainly reported under LF1M and ML1M.
The algorithm is especially effective under ML1M, where $\Delta$ was reduced in all settings, regardless of the sampling type.
{\color{black} Nonetheless, these results might also derive from the prior unfairness level exhibited by the \emph{Base} models under LF1M and ML1M.
Specifically, compared with the smaller datasets FNYC and FTKY, the prior gap in recommendation utility is much more pronounced under LF1M and ML1M, which clearly enables the augmented graph to make a positive impact to a greater extent.
}

\begin{figure}
    \centering
    \setlength{\tabcolsep}{15pt}
    \resizebox{1.02\linewidth}{!}{
    \fontsize{36pt}{36pt}\selectfont
    \begin{tabular}{cc}
        \multicolumn{2}{c}{\includegraphics[width=\textwidth]{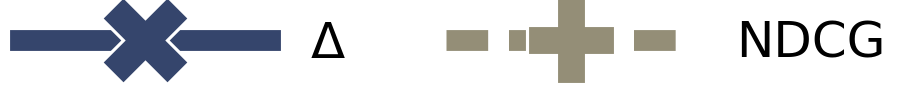}} \\[1pt]
        \textbf{FR + PR | NCL | LF1M | Gender |} $\mathbf{\Psi_{\mathcal{I}} = 20\%}$ & \textbf{FR + PR | NCL | LF1M | Gender |} $\mathbf{\Psi_{\mathcal{U}} = 35\%}$ \\
        \includegraphics{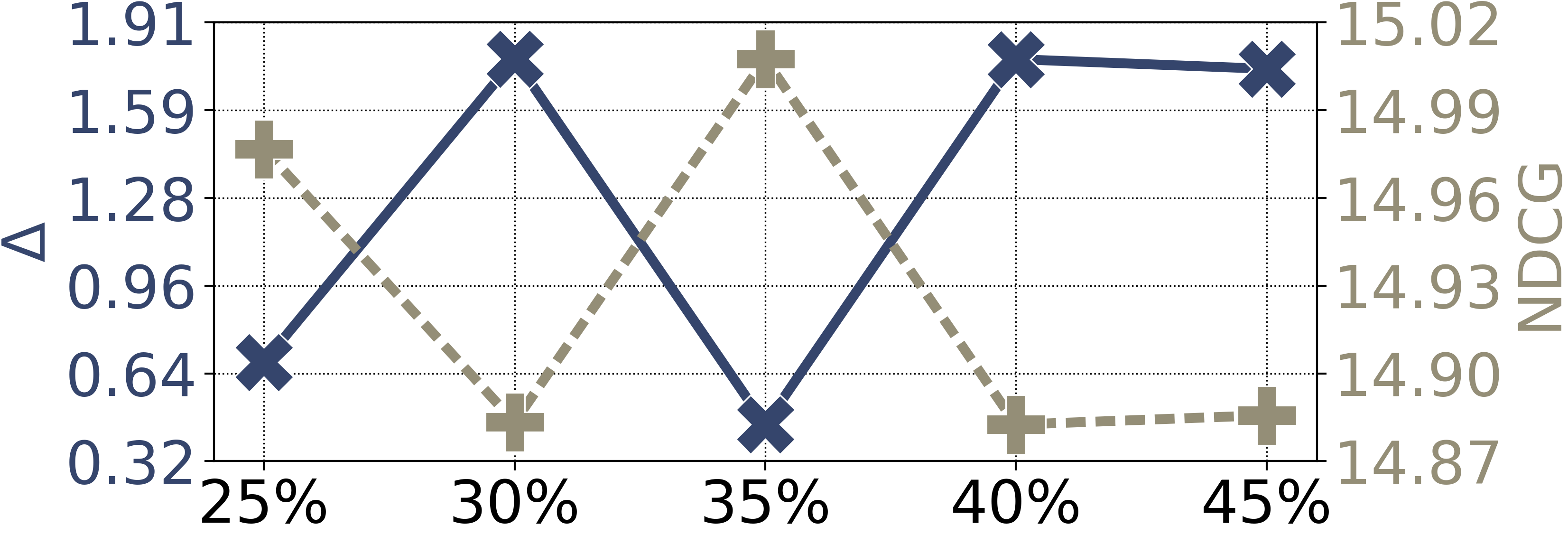} & \includegraphics{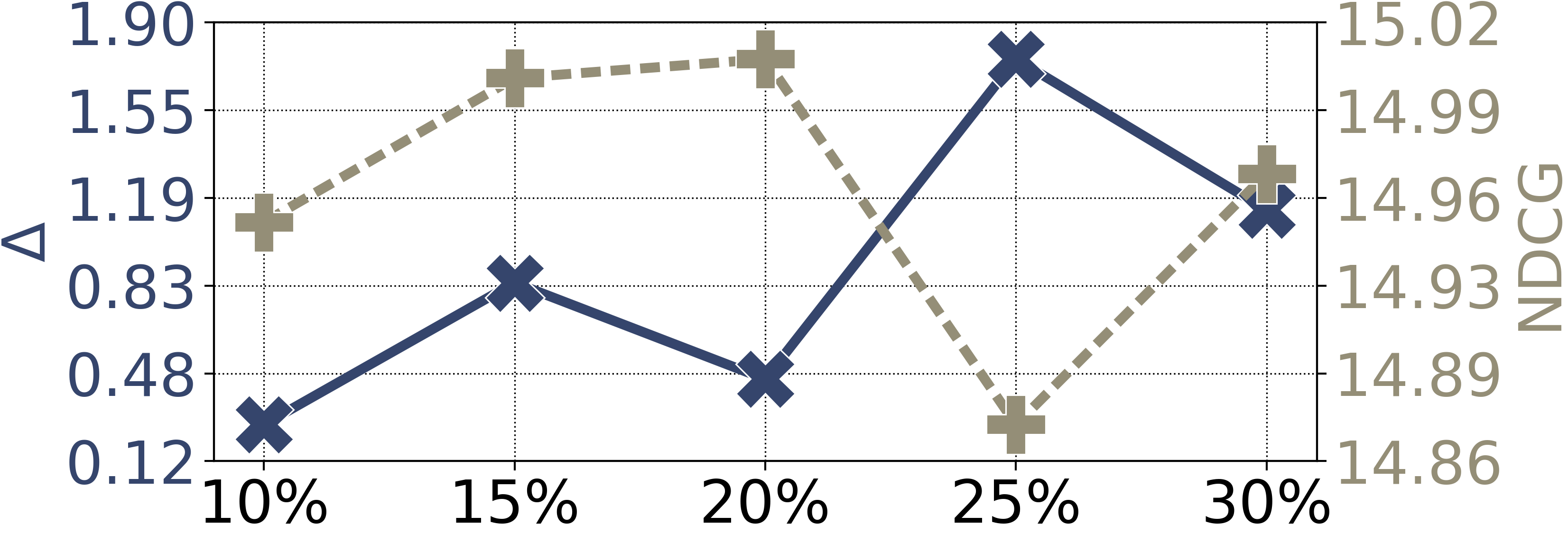} \\
        $\Psi_{\mathcal{U}}$ & $\Psi_{\mathcal{I}}$ \\[20pt]
        \textbf{IR + IP | SGL | ML1M | Gender |} $\mathbf{\Psi_{\mathcal{I}} = 20\%}$ & \textbf{IR + IP | SGL | ML1M | Gender |} $\mathbf{\Psi_{\mathcal{U}} = 35\%}$ \\
        \includegraphics{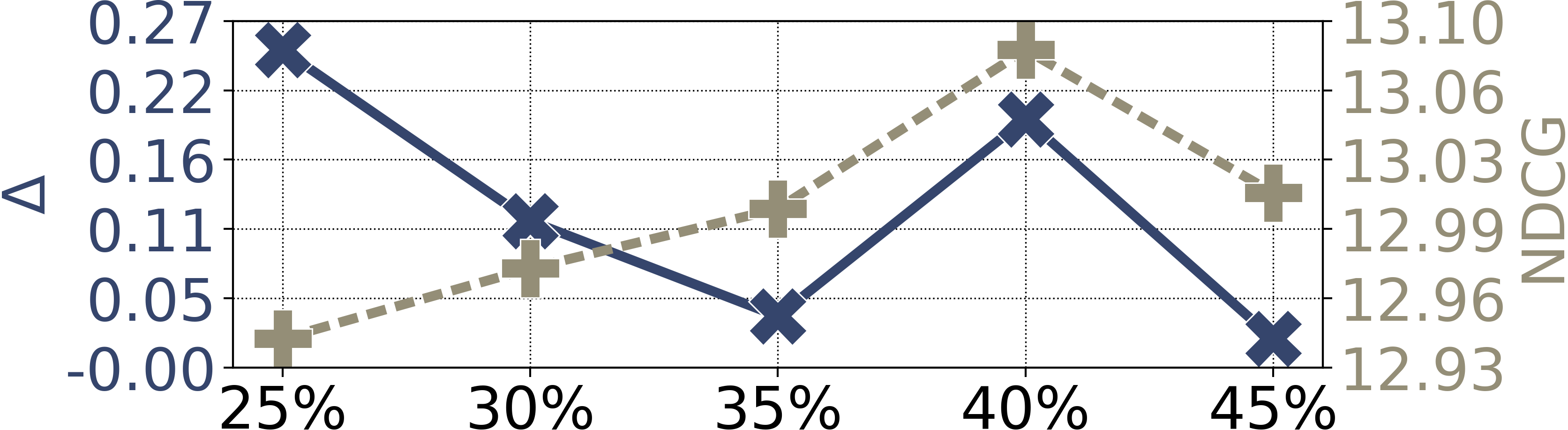} & \includegraphics{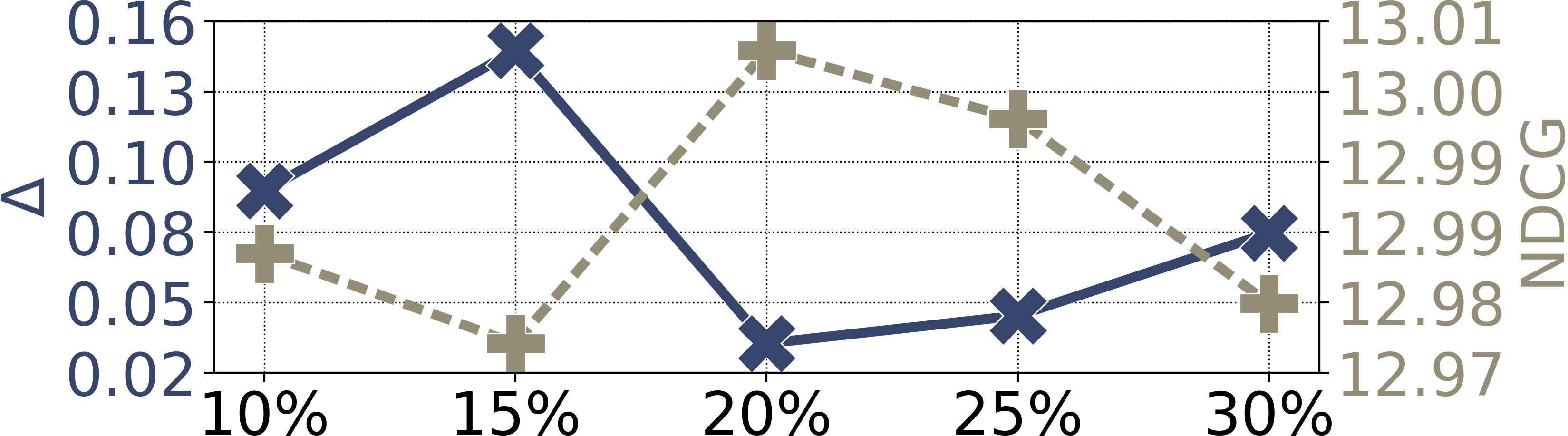} \\
        $\Psi_{\mathcal{U}}$ & $\Psi_{\mathcal{I}}$ \\
    \end{tabular}
    }
    \caption{Impact of $\Psi_{\mathcal{U}}$ and $\Psi_{\mathcal{I}}$ on fairness ($\Delta$) and utility (NDCG) levels on the settings identified by each subplot title.}
    \Description{Lineplots that display the change in recommendation utility and fairness as the sampling policies parameters $\Psi_{U}$ and $\Psi_{I}$ vary.}
    \label{fig:policy_psi_impact}
\end{figure}

Results under FNYC also exhibit two outliers (PR on HMLET and ZN+IP on LightGCN) with a $\Delta$ level significantly higher than the \emph{Base} one.
These under-performing settings are the result of the epoch selection process adopted by the reproduced augmentation technique.
In detail, as soon as some edges are added to the graph, unfairness levels are recorded and the augmentation process continues until the early stopping criterion triggers.
The augmentation step resulting in the lowest $\Delta$ is considered the fairest, even if this $\Delta$ is notably higher than the starting one, which can lead to potential outliers as PR on HMLET and ZN+IP on LightGCN under FNYC.

Under LF1M, IP on HMLET and FR on LightGCN report the highest level of unfairness mitigation, even in combination with less effective policies, e.g., IR+IP on HMLET.
This observation suggests the models potentially learnt unfair signals from different clustered structures in the graph.
Precisely, the considered policies positively affected disadvantaged users with similar preferences (IP) and those with tastes dissimilar to advantaged users (FR).
FR exhibited excellent mitigation performance also under ML1M on both models.
Therefore, we can state that the algorithm is effective in mitigating unfairness when the augmentation targets \emph{disadvantaged users who are furthest from advantaged users} in the graph.
Although this work focuses only on applying the method for unfairness mitigation, such an observation can be used to provide a user-friendly explanation of the disparity in NDCG across gender groups, as remarked in \cite{DBLP:conf/cikm/BorattoFFMM23}.

\begin{table*}
    \centering
    \caption{Utility (NDCG) and fairness ($\Delta$) percentage levels before (\emph{Base}) and after (\emph{Aug}) adopting the augmented graph on non-augmentable GNNs and non-GNN models. Underlined values denote the augmented graph led to improvements as NDCG increments and $\Delta$ decrements. Best NDCG and $\Delta$ values in each row are in bold. \emph{G} stands for \emph{Gender}, \emph{A} for \emph{Age}.}
    \resizebox{\textwidth}{!}{
    \begin{tabular}{lcll|rr|rr|rr|rr|rr|rr|rr}
    \toprule
      &  &  &  & \multicolumn{4}{c|}{Weak Transferability} & \multicolumn{10}{c}{Strong Transferability} \\
      & Aug. &  &  & \multicolumn{2}{c|}{UltraGCN} & \multicolumn{2}{c|}{SVD-GCN-S} & \multicolumn{2}{c|}{BPR} & \multicolumn{2}{c|}{NeuMF} & \multicolumn{2}{c|}{ENMF} & \multicolumn{2}{c|}{EASE} & \multicolumn{2}{c}{DiffRec} \\
     & Graph &  &  & NDCG $\uparrow$ & $\Delta$ $\downarrow_0$ & NDCG $\uparrow$ & $\Delta$ $\downarrow_0$ & NDCG $\uparrow$ & $\Delta$ $\downarrow_0$ & NDCG $\uparrow$ & $\Delta$ $\downarrow_0$ & NDCG $\uparrow$ & $\Delta$ $\downarrow_0$ & NDCG $\uparrow$ & $\Delta$ $\downarrow_0$ & NDCG $\uparrow$ & $\Delta$ $\downarrow_0$ \\
    \midrule
    \multirow[c]{2}{*}{FNYC} & AutoCF & \multirow[c]{2}{*}{G} & Base & 7.80 & 0.10 & 4.24 & 0.33 & 7.75 & 0.09 & 8.06 & 0.11 & 0.90 & \textbf{0.03} & 8.41 & 0.34 & \textbf{8.43} & 0.25 \\
     & ZN+IT &  & Aug & 7.55 & \underline{\textbf{0.02}} & 4.02 & 0.34 & 7.69 & 0.21 & 7.76 & 0.24 & 0.77 & 0.06 & 8.03 & 0.62 & \textbf{8.22} & \underline{0.13} \\
    \cline{1-18}
    \multirow[c]{2}{*}{FTKY} & LightGCN & \multirow[c]{2}{*}{G} & Base & 9.54 & \textbf{0.03} & 5.41 & 0.96 & 9.35 & 0.16 & 9.50 & 0.09 & 1.84 & 0.44 & 9.72 & 0.15 & \textbf{10.06} & 0.17 \\
     & FR & & Aug & 9.47 & 0.10 & \underline{5.44} & 0.96 & 9.18 & \underline{\textbf{0.04}} & 9.47 & \underline{0.08} & 1.67 & \underline{0.28} & 9.67 & 0.19 & \textbf{9.93} & 0.19 \\
    \cline{1-18}
    \multirow[c]{2}{*}{LF1M} & XSimGCL & \multirow[c]{2}{*}{G} & Base & 18.77 & 3.00 & 15.97 & 2.78 & 18.42 & 3.78 & 17.92 & 3.37 & 1.00 & \textbf{0.21} & \textbf{22.81} & 2.56 & 21.63 & 2.70 \\
     & ZN+PR & & Aug & 18.37 & 4.13 & 15.86 & 3.49 & 18.16 & 4.12 & 17.28 & 3.55 & 0.96 & \underline{\textbf{0.14}} & \textbf{22.78} & 2.65 & 21.73 & 3.19 \\
    \cline{1-18}
     \multirow[c]{2}{*}{ML1M} & SGL & \multirow[c]{2}{*}{G} & Base & 11.94 & 1.80 & 6.14 & \textbf{0.56} & 12.22 & 1.73 & 11.87 & 1.84 & 2.70 & 0.63 & \textbf{12.63} & 1.90 & 12.60 & 2.21 \\
     & IR+IP & & Aug & \underline{11.99} & 2.07 & \underline{6.24} & 1.19 & 11.62 & 2.16 & 11.82 & 1.84 & 2.62 & \textbf{0.83} & \textbf{12.62} & 1.97 & 12.58 & 2.33 \\
    \cline{1-18}
    \multirow[c]{2}{*}{RENT} & XSimGCL & \multirow[c]{2}{*}{A} & Base & 0.62 & 0.38 & \textbf{1.01} & 0.29 & 0.58 & 0.32 & 0.62 & 0.32 & 0.23 & \textbf{0.06} & 0.98 & 0.77 & 0.64 & 0.22 \\
     & ZN+PR & & Aug & 0.44 & \underline{0.34} & \textbf{1.01} & 0.33 & 0.58 & \underline{0.25} & \underline{0.70} & \underline{0.29} & 0.10 & \textbf{0.06} & 0.97 & \underline{0.75} & 0.55 & 0.37 \\
    \bottomrule
    \end{tabular}
    }
    \label{tab:transfer_learning}
\end{table*}

\subsection{RQ3: Impact of $\Psi$ on Sampling Policies}

We observed that targeting specific nodes via sampling policies enhances the augmentation algorithm effectiveness in mitigating unfairness and increasing utility.
However, such policies are driven by crucial parameters to determine whether a node will be targeted by the augmentation process.
To this end, we address an aspect that was not covered in \cite{DBLP:conf/cikm/BorattoFFMM23}, namely, the analysis of the impact of $\Psi_{\mathcal{U}}$ and $\Psi_{\mathcal{I}}$ on the resulting utility and fairness levels.
To assess their variation across different sampling policies, we focus on settings that reflected the power of the augmentation process the most.
According to Table~\ref{tab:overall_performance}, FR+PR on NCL under LF1M across gender groups and IR+IP on SGL under ML1M across gender groups are two settings under the largest datasets where the mitigation procedure led to significant unfairness mitigation levels and the overall utility was also increased.
We evaluate these policies on the selected settings by varying $\Psi_{\mathcal{U}}$ from the set $\{25\%, 30\%, 35\%, 40\%, 45\%\}$ and $\Psi_{\mathcal{I}}$ from $\{10\%, 15\%, 20\%, 25\%, 30\%\}$.
We vary only one parameter at a time to isolate the impact of each policy type.

Figure~\ref{fig:policy_psi_impact} reports the impact of varying $\Psi_{\mathcal{U}}$ and $\Psi_{\mathcal{I}}$ on the fairness and utility levels.
Interestingly, the $\Psi_{\mathcal{U}}$ and $\Psi_{\mathcal{I}}$ values reported in \cite{DBLP:conf/cikm/BorattoFFMM23} turn out to be among the most effective combinations across the selected settings.
We can consider the gap between the NDCG and $\Delta$ lines as a representation of the trade-off between fairness and utility.
Hence, it is evident that the combination $\langle\Psi_{\mathcal{U}} = 35\%$, $\Psi_{\mathcal{I}} = 20\%\rangle$ outlines a good trade-off across several settings.
For instance, setting $\Psi_{\mathcal{U}}$ to $45\%$ results in a better trade-off than $35\%$ on SGL, but not on NCL.
Similarly, setting $\Psi_{\mathcal{I}}$ to $10\%$ reports the lowest $\Delta$ on NCL, but it under-performs on SGL.

Except when $\Psi_{\mathcal{U}}$ varies on SGL, the subplots delineate a consistent trend where reducing $\Delta$ causes the NDCG to increase, and vice versa.
This trend of inverse proportion is especially prominent on the scenarios where $\Psi_{\mathcal{I}}$ varies.
For instance, setting $\Psi_{\mathcal{I}}$ to $15\%$ ($25\%$) results in the maximum disruption in fairness and utility levels, but increasing (decreasing) the value to $20\%$ completely reverses the trend on SGL (NCL).
This observation might be influenced by the augmentation logic itself: new edges are specifically identified to increase the utility of the disadvantaged group such that it matches that of the advantaged one (resulting in fairer recommendations).
It follows that selecting edges that negatively influence the utility of $\mathcal{U}_D$ potentially results in $\Delta$ being negatively affected as well.

Most of the subplots in Figure~\ref{fig:policy_psi_impact} reports the extreme values as the ones leading to the worst trade-offs between utility and fairness.
Especially selecting a smaller sample of users or items negatively affects both utility and fairness.
This trend might be caused by the limited augmentable portion of the graph, which could critically exclude valuable nodes.
Yet, less favorable combinations of $\Psi_{\mathcal{U}}$ and $\Psi_{\mathcal{I}}$ still result in augmented graphs that produce fairer recommendations compared with the \emph{Base} model.
The interested reader could compare Figure~\ref{fig:policy_psi_impact} with Table~\ref{tab:overall_performance} to confirm such statement.

As a summary, adhering to the parameters suggested in \cite{DBLP:conf/cikm/BorattoFFMM23} ($\Psi_{\mathcal{U}} = 35\%$ and $\Psi_{\mathcal{I}} = 20\%$) ensures an augmentation process that maintains a favorable trade-off across diverse datasets and models.

\subsection{RQ4: Fair Augmentation Transfer Learning} \label{subsec:re-training}

As previously anticipated, some GNNs and any non-GNN model are not augmentable.
Therefore, to evaluate the effectiveness of the reproduced mitigation method on such systems, we explore the potentiality of augmented graphs to generate fairer recommendations when used to re-train non-augmentable models.
This can be viewed as a transfer learning process, where the fair knowledge embedded in augmented graphs is transferred to another model.
We distinguish between \emph{weak transferability}, when the re-trained model is a non-augmentable GNN, and \emph{strong transferability}, when the re-trained model is not a GNN.
The term transferability slightly refers to the homonymous property proposed in \cite{DBLP:journals/ipm/BorattoFMM23}, given that it concerns models for which the augmentation process is not applicable.

We selected a single setting for each dataset to re-train the non-augmentable models introduced in Section~\ref{subsec:exp-setting}.
Specifically, the experiments in Table~\ref{tab:overall_performance} reporting the best trade-off between fairness and utility were taken as the candidate augmented graphs for re-training, prioritizing those where unfairness was mitigated the most.
We selected AutoCF under FNYC, LightGCN under FTKY, XSimGCL under LF1M across gender groups, SGL under ML1M across gender groups, XSimGCL under RENT.
Due to the augmentation process being ineffective under RENT, we opted for the setting exhibiting the highest utility.
We expect the models re-trained on this setting to report unfair recommendations as well, so we treat this setting as a control result to validate the experiment.

Table~\ref{tab:transfer_learning} reports the impact of the selected augmented graphs on the utility (NDCG) and fairness ($\Delta$) levels of the tested non-augmentable GNNs and non-GNN models.
It is evident that the fair and high-utility knowledge embedded in the augmented graphs was not transferred as expected.
{\color{black} A few improvements were reported to a greater extent under RENT and FTKY.
Nonetheless, as emphasized in Section \ref{sec:comparison_across_policies}, the prior unfairness level of the \emph{Base} models under these datasets is low.
Hence, these improvements are minimal and not significant.
In line with these observations}, the overall impact of the augmented graphs is negligible and not consistent.

Across transferability types, it can be noticed that the non-augmentable GNNs were slightly influenced by the augmented graphs, regardless the effect being positive or negative.
On the other hand, applying the augmented graphs for strong transferability led to varied outcomes across the models.
However, the augmented graphs mostly led to negative outcomes compared with the \emph{Base} one, e.g., utility reductions on BPR under ML1M and unfairer recommendations on DiffRec under LF1M.

Interestingly, EASE under LF1M and DiffRec under FNYC and FTKY report the highest overall utility under the corresponding datasets, even in comparison with the GNNs in Table~\ref{tab:overall_performance}.
Nonetheless, the augmentation process enables XSimGCL to generate fairer recommendations than EASE under LF1M and surpass DiffRec in utility under FNYC and FTKY.
{\color{black} This observation underscores i) the benefits that can be offered by graph augmentations, and ii) the challenging task of mitigating unfairness and bias from pre- and in-processing stages.
The former point is remarkable in the study of \cite{DBLP:conf/cikm/BorattoFFMM23} because the utility increment goal was not directly pursued, though it being implicitly addressed in their augmentation process that targets the disadvantaged group.
The latter points deals with the intricate challenge of counteracting unfairness issues in recommendation by just manipulating the data or the model learning process, as also emphasized by recent studies~\cite{DBLP:journals/ipm/BorattoFMM23}.}

Overall, the fair augmented graph did not exhibit an effective transferability across models, regardless they being GNNs or not.

\section{Conclusions} \label{sec:conclusions}

We conducted a reproducibility study that not only assessed the effectiveness of the reproduced algorithm in mitigating consumer unfairness in GCF, but also addressed significant shortcomings in the original work methodology.
In detail, Section~\ref{subsec:policy_formalization} formally defined the policies used to restrict the user and item set, Section~\ref{subsec:aug_formalization} provided a formalization of the reproduced augmentation pipeline and its impact on the prediction step of GNN-based models, Section~\ref{subsec:extended_policies} presented a novel set of policies that also contemplate classical graph properties as pagerank and the time feature available in several public datasets used in recommendation.
{\color{black} Section~\ref{subsec:re-training} explored the transferability of the fair knowledge instilled in the augmented graph during the re-training stage of non-augmentable GNNs and non-GNNs recommender systems.}

Our comprehensive experimental evaluation shed light on key aspects of the reproduced mitigation procedure: (i) its performance in mitigating unfairness and in increasing utility improves as the dataset size increases, (ii) it is effective especially when applied with the policies including FR {\color{black} due to the isolated clustering structure captured by such a policy}, (iii) its resulting levels of utility and fairness exhibit a trend of inverse proportion as $\Psi_{\mathcal{U}}$ or $\Psi_{\mathcal{I}}$ varies, (iv) the fair and high-utility knowledge embedded in the augmented graph is not transferable to other models.
Additionally, we offer an updated framework, referred to as FA4GCF, that relies on efficient GNN backbones and integrates a simple interface to seamlessly adapt the augmentation process to diverse settings in terms of models, datasets, and sampling policies.
Our findings highlight the relevance of our contributions with respect to the original work.

We acknowledge two limitations of our study.
The former regards an emphasized issue in the literature of consumer fairness~\cite{DBLP:journals/ipm/BorattoFMM23}, that is the limited datasets with user demographic information, which causes the corresponding research to focus on just a few datasets~\cite{DBLP:journals/air/VassoyL24}.
Although we selected 5 datasets across 4 diverse topics, larger datasets with sensitive attributes are publicly available, e.g., Last.FM-2b~\cite{DBLP:journals/ipm/MelchiorreRPBLS21}, Alibaba~\cite{DBLP:conf/www/PeiYCLSJOZ19}.
However, concerns about incomplete information, e.g., lack of users' identifiers in Alibaba, or challenging computational resources arise when working with such corpora.
{\color{black} Indeed, research into the consumer fairness on the interaction data of the Last.FM platform typically leverages samples of Last.FM-2b due to its remarkable size~\cite{DBLP:conf/ecir/BalloccuBCFM23,DBLP:journals/ipm/BorattoFMM23}, while fairness studies on the full corpus are limited~\cite{DBLP:journals/ipm/MelchiorreRPBLS21}.}
The latter limitation regards the extension of the weak transferability approach to other augmentable GNNs.
UltraGCN and SVD-GCN-S are two models of a small class of GNNs that perform an implicit message passing in GCF.
Therefore, it is still open the doubt that the fair knowledge in the augmented graph could be transferred across GNNs, regardless of them being augmentable or not.
Such observation gives rise to potential future works, {\color{black} such as exploring graph augmentations that might transfer the fair knowledge universally to any GNNs or to a restricted family, e.g., GNNs based on contrastive learning}.

Other lines of future research will study novel techniques to achieve a successful transferability across different recommender systems or to extend the set of augmentable models.
{\color{black} For instance, adding a simple graph transformation to the predictions process of non-augmentable systems would enable such models to generate recommendations from the resulting graph embedding layers and, hence, learn a fair augmentation from the reproduced approach.}
Moreover, the augmentation process is also presented by \cite{DBLP:conf/cikm/BorattoFFMM23} as a tool to generate explanations of consumer unfairness.
Nevertheless, the application of the approach for such purposes was not deeply explored and still remains an open question.

\begin{acks}
We acknowledge financial support from (i) the National Recovery and Resilience Plan (NRRP), Mission 4 Component 2 Investment 1.1 - Call for tender No. 3277, published on December 30, 2021, by the Italian Ministry of University and Research (MUR), funded by the European Union – Next Generation EU. Project Code ECS0000038 – Project Title eINS Ecosystem of Innovation for Next Generation Sardinia – Grant Assignment Decree No. 1056 adopted on June 23, 2022, by the MUR (CUP F53C22000430001) and (ii) the project PHaSE - Promoting Healthy and Sustainable Eating through Interactive and Explainable AI Methods, funded by MUR under the PRIN 2022 program (CUP H53D23003530006).
\end{acks}

\balance
\bibliographystyle{ACM-Reference-Format}
\bibliography{sample-base}

\end{document}